\documentclass[prb,showpacs,twocolumn,floatfix]{revtex4} 
\usepackage{amsmath}
\usepackage{graphicx}
\usepackage{float}
\usepackage{amssymb}
\usepackage{times}
\usepackage[colorlinks,hyperindex]{hyperref}
\DeclareGraphicsExtensions{.eps,.ps}
\hypersetup
	{
		colorlinks,%
		citecolor=black,%
		linkcolor=blue,
		urlcolor=black,%
	}

\makeatletter

\newcommand{\Rmnum}[1]{\expandafter\@slowromancap\romannumeral #1@}
\newcommand{\be}{\begin{equation}}
\newcommand{\ee}{\end{equation}}
\newcommand{\beq}{\begin{eqnarray}}
\newcommand{\eeq}{\end{eqnarray}}
\makeatother

\begin{document}
\title{Quasi-adiabatic quantum Monte Carlo algorithm for quantum evolution in imaginary time}

\author{Cheng-Wei Liu}

\author{Anatoli Polkovnikov}

\author{Anders W. Sandvik}
\affiliation{Department of Physics, Boston University, 590 Commonwealth Avenue, Boston, Massachusetts 02215, USA}

\date{\today}

\begin{abstract}
We propose a quantum Monte Carlo (QMC) algorithm for non-equilibrium dynamics in a system with a parameter varying as a function of imaginary time. The method 
is based on successive applications of an evolving Hamiltonian to an initial state  and delivers results for a whole range of the tuning parameter 
in a single run, allowing for access to both static and dynamic properties of the system. This approach reduces to the standard Schr\"odinger dynamics 
in imaginary time for quasi-adiabatic evolutions, i.e., including the leading non-adiabatic correction to the adiabatic limit.  
We here demonstrate this {\it quasi-adiabatic QMC} (QAQMC) method for linear ramps of the transverse-field Ising model across its  quantum-critical point in 
one and two dimensions. The critical behavior can be described by generalized dynamic scaling. For the two-dimensional square-lattice system we use 
the method to obtain a high-precision estimate of the quantum-critical point $(h/J)_c=3.04458(7)$, where $h$ is the transverse magnetic field and 
$J$ the nearest-neighbor Ising coupling. The QAQMC method can also be used to extract the Berry curvature and the metric tensor.

\end{abstract}
\maketitle

\section{Introduction}
\label{intro}

Quantum Monte Carlo (QMC) methods \cite{assaad07,sandvik10} have become indispensable tools for ground-state and finite-temperature studies of many classes 
of interacting quantum systems, in particular those for which the infamous ``sign problem'' can be circumvented.\cite{kaul12} In ground-state projector methods, 
an operator $P(\beta)$ is applied to a ``trial state'' $|\Psi_0\rangle$, such that $|\Psi_\beta\rangle = P(\beta)|\Psi_0\rangle$ approaches  the ground state of the 
Hamiltonian $\mathcal{H}$ when $\beta \to \infty$ and an expectation value  $\langle A\rangle = \langle \Psi_\beta|A|\Psi_\beta\rangle/Z$, with the norm 
$Z = \langle \Psi_\beta|\Psi_\beta\rangle$, approaches its true ground-state value, $\langle A\rangle \to \langle 0| A|0\rangle$. For the projector, one 
can use $P(\beta)=\exp{(-\beta \mathcal{H})}$ or a high power of the Hamiltonian\cite{betanote}, $P(M)=(-\mathcal{H})^M$. Here we will discuss a 
modification of the latter projector for studies of dynamical properties of systems out of equilibrium.

Real-time dynamics for interacting quantum systems is difficult to deal with computationally. Solving the Schr\"odinger equation directly, computations
are restricted to very small system sizes by the limits of exact diagonalization.  Despite progress with the Density-Matrix Renormalization Group (DMRG) 
\cite{PhysRevLett.69.2863, RevModPhys.77.259} and related methods based on matrix-product states, this approach is in practice limited to 
one-dimensional systems and relatively short times. Efficiently studying long-time dynamics of generic interacting quantum systems in higher dimensions 
is still an elusive goal. However, recently, in Ref.~[\onlinecite{degrandi11}], it was demonstrated that real-time and imaginary-time dynamics bear 
considerable similarities, and in the latter case, powerful and high-precision QMC calculations can be carried out on large system sizes for the 
class of systems where sign problems can be avoided.

Our work reported here is a further development of the method introduced in Ref.~[\onlinecite{degrandi11}], where it was realized that a modification  
of the ground-state projector Monte Carlo approach with $P(\beta)=\exp{(-\beta \mathcal{H})}$ can be used to study non-equilibrium set-ups in quantum 
quenches (or ramps), where  a parameter of the Hamiltonian depends on time according to an arbitrary protocol. By performing a standard Wick rotation 
of the time axis,  a wave function is governed 
by the Shr\"odinger equation in imaginary time $t=-i\tau$ ($\tau$ being real),
\begin{equation}
\partial_\tau |\psi (\tau)\rangle = -\mathcal{H}[\lambda(\tau)]|\psi (\tau)\rangle.
\label{schrod}
\end{equation}
Here the Hamiltonian depends on the parameter 
$\lambda$ through time, e.g., 
\begin{equation}
\label{eq:h}
\mathcal{H}=\mathcal{H}_0 + \lambda(\tau)V, 
\end{equation}
where $V$ and $H_0$ typically do not commute. The method is not limited to this form, however, and any evolution of $\mathcal{H}$ can be considered.
The Schr\"odinger equation has the formal solution 
\begin{equation}
|\psi(\tau)\rangle=U(\tau)|\psi(\tau_0)\rangle, 
\label{psitauformal}
\end{equation}
where the imaginary-time evolution operator is given by
\begin{equation}
U(\tau)=T_\tau {\rm exp}\left [ - \int_{\tau_0}^\tau d \tau' \mathcal{H}[\lambda(\tau')] \right ],
\label{utau}
\end{equation}
where $T_\tau$ indicates time ordering. A time-evolved state $U(\tau)|\Psi(\tau_0)\rangle$ and associated expectation values can be sampled using a 
generalized projector QMC algorithm. In Ref.~[\onlinecite{degrandi11}] it was demonstrated that this non-equilibrium QMC (NEQMC) approach can be applied to 
study dynamic scaling at quantum phase transitions, and there are many other potential applications as well, e.g., when going beyond studies of  finite-size 
gaps in ``glassy'' quantum dynamics and the quantum-adiabatic paradigm for quantum computing. 

Here we introduce a different approach to QMC studies of quantum quenches, which gives results for a whole range of parameters 
$\lambda \in [\lambda(\tau_0),\lambda(\tau)]$ in a single run (instead of just the final time), at a computational effort comparable to 
the previous approach. Instead of using the conventional time-evolution operator Eq.~(\ref{utau}), we consider a generalization of the equilibrium 
QMC scheme based on projection with $(-\mathcal{H})^M$, acting on the initial ground state of $\mathcal{H}[\lambda(\tau_0)]$ with a product
of evolving Hamiltonians:
\begin{equation}
P_{M,1}=[-\mathcal{H}(\lambda_M)]....[-\mathcal{H}(\lambda_2)][-\mathcal{H}(\lambda_1)],
\label{p1m}
\end{equation}
where 
\begin{equation}
\lambda_t= \lambda_0+t\Delta_\lambda,
\label{lambdat}
\end{equation}
and $\Delta_\lambda=[\lambda_{M}-\lambda_0]/M$ is the single-step change in the tuning parameter.\cite{gridnote} Here we will consider a 
case where the ground state $|\Psi(\lambda_0)\rangle$ of $\mathcal{H}(\lambda_0)$ is known and easy to generate (stochastically or otherwise) and 
the ground states for other $\lambda$-values of interest are non-trivial. The stochastic sampling used to compute the evolution then takes place 
in a space representing path-integral-like terms contributing to the matrix element (the norm) $\langle \Psi(\lambda_0)| P_{1,M} P_{M,1}|\Psi(\lambda_0)\rangle$. 
We will also later consider a modification of the method in which the ground state at the final point $\lambda_M$ is known as well, in which case contributions 
to $\langle \Psi(\lambda_M)| P_{M,1}|\Psi(\lambda_0)\rangle$ are sampled. 

Staying with the doubly-evolved situation for now, we evaluate generalized expectation values 
after $t$ out of the $M$ operators in the product (\ref{p1m}) have acted:
\begin{equation}
\langle A\rangle_t = \frac{\langle \Psi(\lambda_0)| P_{1,M} P_{M,t+1}AP_{t,1}|\Psi(\lambda_0)\rangle}{\langle \Psi(\lambda_0)| P_{1,M} P_{M,1}|\Psi(\lambda_0)\rangle}.
\label{atdef}
\end{equation}
We will refer to this matrix element as an {\it asymmetric expectation value}, with the special case $t=M$ corresponding to a true quantum-mechanical 
expectation value taken with respect to an evolved wave function, 
\begin{equation}
\label{evolved_wf}
|\psi_M\rangle =\frac{P_{M,1}|\Psi(\lambda_0)\rangle}{\sqrt{\langle \Psi(\lambda_0)| P_{1,M} P_{M,1}|\Psi(\lambda_0)\rangle} },
\end{equation}
which approaches the ground state $|\Psi[\lambda(\tau_M)]\rangle$ of the Hamiltonian $H[\lambda(\tau_M)]$ for $M \to \infty$. 

Away from the adiabatic limit, the evolved wave function Eq.~(\ref{evolved_wf}) is, generally speaking, not the ground state of the equilibrium system. Nevertheless,  
as we demonstrate in detail in Sec.~\ref{sec:apt}, a quench velocity $v \propto \Delta_\lambda N$ can be defined such that the symmetric expectation value 
$\langle A\rangle_{t=M}$ in Eq.~(\ref{atdef}) 
approaches the expectation value $\langle A(\tau=t)\rangle$ after a conventional linear imaginary-time quantum quench with Eq.~(\ref{utau}) done with the same velocity 
$v$, if $v$ is low enough. In fact, the two quantities are the same to leading (linear) order in $v$, not only in the strict adiabatic limit $v \to 0$. We therefore name 
this scheme the {\it quasi-adiabatic} QMC (QAQMC) algorithm.  Importantly, the leading corrections to the adiabatic evolution of the asymmetric expectation values for any $t$ 
contain important information about non-equal time correlation functions, very similar to the imaginary-time evolution. 

The principal advantage of QAQMC  over the NEQMC approach 
is that expectation values of diagonal operators in the basis used can be obtained simultaneously for the  whole evolution path $\lambda_0\ldots \lambda_M$, by measuring 
$\langle A\rangle_t$ in Eq.~(\ref{atdef}) at arbitrary $t$ points \cite{averagenote} (and one can also extend this to general off-diagonal operators, along the lines of Ref.~[\onlinecite{dorneich01}], but we here limit studies to diagonal operators). The QAQMC scheme is also easier to implement in practice than the NEQMC method because there are no time integrals to sample.

As mentioned above, we will here have in mind a situation where the initial state $|\Psi(\lambda_0)\rangle$ is in some practical sense ``simple," but this is 
not necessary for the method to work---any state that can be simulated with standard equilibrium QMC methods can be used as the initial state for the dynamical 
evolution. The final evolved state $|\psi_M\rangle$ can be very complex, e.g., for a system in the vicinity of a quantum-critical point or in a 
``quantum glass'' (loosely speaking, a system with slow intrinsic dynamics due to spatial disorder and frustration effects). Here, as a demonstration of 
the correctness and utility of the QAQMC approach, we study generalized dynamic scaling in the neighborhood of the quantum phase transitions in the standard 
one-dimensional (1D) and 2D transverse-field Ising models (TFIMs).

As noted first in Ref.~[\onlinecite{degrandi11}], the NEQMC method can be used to extract the components of the quantum metric tensor,\cite{provost_80} the diagonal elements 
of which are the more familiar fidelity susceptibilities. Thanks to its ability to capture the leading non-adiabatic corrections to physical observables, 
the QAQMC approach can also be used for this purpose, and, as we will discuss briefly here and in more detail in Ref.~[\onlinecite{adi_long}], one can also 
extract the Berry curvature through the imaginary antisymmetric components of the geometric tensor

The rest of the paper is organized in the following way. In Sec.~\ref{sec:apt}, we use adiabatic perturbation theory (APT) to demonstrate the ability of the 
QAQMC scheme to correctly capture the standard Schr\"odinger evolution in imaginary time, not only in the adiabatic limit but also including  the leading 
corrections in the quench velocity. We show how these leading corrections correspond to the geometric tensor. In Sec.~\ref{sec:results}, we discuss tests of 
the QAQMC scheme on 1D and 2D TFIMs, and also present a high-precision result for the critical field in the 2D model. In Sec.~\ref{sec:conclusions},  
we summarize our main  conclusions and discuss future potential applications of the algorithm.

\section{Adiabatic perturbation theory}
\label{sec:apt}

The key question we address in this section is whether the matrix element $\langle A\rangle_t$ in Eq.~(\ref{atdef}) can give useful dynamical information 
for arbitrary ``time'' points $t$ in the sequence of $2M$ operators. The expression only reduces to a conventional expectation value at the symmetric point 
$t=M$, and even there it is not clear from the outset how $\langle A\rangle_{t=M}$ computed for different $M$ relates to the velocity  dependence of the 
expectation value $\langle \Psi (0)|U^{*}(\tau) \hspace{1pt} A \hspace{1pt} U(\tau)|\Psi(0)\rangle$ based on the Schr\"odinger time-evolution operator  in Eq.~(\ref{utau}). Going away 
from the symmetric point brings in further issues to be addressed. For instance, there is no variational property of the asymmetric expectation value 
$\langle \mathcal{H} \rangle_t$ of the Hamiltonian for $t\not=M$.  Nevertheless, the approach to the adiabatic limit is well behaved and we can associate 
the leading deviations from adiabaticity with well defined dynamical correlation functions that appear as physical response in real time protocols. We show 
here, for the linear evolution Eq.~(\ref{lambdat}), that one can identify a velocity $v \propto N/M$ such that a linear imaginary-time quench with $\lambda_t =vt$ 
in Eq.~(\ref{lambdat}) gives the same  results in the two approaches when $t=M$, including the leading (linear) corrections in $v$. For $t\not=M$, the relevant 
susceptibilities in QAQMC defining non-adiabatic response are different than at $t=M$ but still well defined, contain useful information, and obey generic scaling 
properties.  

In order to facilitate the discussion of the QAQMC method, we here first review the previous APT approach for Schr\"odinger imaginary-time dynamics 
\cite{degrandi11, adi_long} and then derive analogous expressions for the  product-evolution. After this, we discuss some properties of the symmetric 
and asymmetric expectation values.

\subsection{Imaginary-time Schr\"odinger dynamics}

The NEQMC method \cite{degrandi11} uses a path-integral-like Monte Carlo sampling to solve the imaginary-time Shcr\"odinger equation Eq.~(\ref{schrod}) for a Hamiltonian $\mathcal{H}[\lambda(\tau)]$ with a time-dependent coupling. The formal solution at time $\tau$ is given by the evolution operator Eq.~(\ref{utau}). 
In the strict adiabatic limit, the system will follow the instantaneous ground state, while in the slow limit one can anticipate deviations from adiabaticity, which will become more severe in gapless systems and, in particular, near phase transitions. Let us discuss the leading non-adiabatic correction to this imaginary-time evolution. The natural way to address this question is to use APT, similar to that developed in Refs.~[\onlinecite{ortiz_2008}] and [\onlinecite{PhysRevB.81.224301}] in real time. We here follow closely the discussion of the generalization to imaginary time in Ref.~[\onlinecite{degrandi11}].

We first write the wave function in the instantaneous eigenbasis $\{ |n(\lambda)\rangle \}$ of the time-dependent Hamiltonian $\mathcal{H}[\lambda(\tau)]$:
\be
| \psi(\tau) \rangle=\sum_n a_n(\tau) |n(\lambda(\tau))\rangle.
\label{eq 1}
\ee
We then substitute this expansion into Eq.~(\ref{schrod}),
\be
{d a_n\over d\tau}+\sum_m a_m(\tau) \langle n|\partial_\tau |m\rangle = -\mathcal E_n (\lambda) \, a_n(\tau),
\label{eq 2}
\ee
where $\mathcal E_n(\lambda)$ are the eigenenergies of the Hamiltonian $\mathcal{H}(\lambda)$ corresponding to the states $|n\rangle$ for
this value of $\lambda$. Making the transformation 
\begin{equation}
a_n(\tau)=\alpha_n(\tau)\exp\left[\int_\tau^0 \mathcal E_n(\tau')d\tau'\right],
\end{equation}
we can rewrite Eq.~(\ref{schrod}) as an integral equation;
\beq
&&\alpha_n(\tau)=\alpha_n(0)+\sum_m \int^0_{\tau} d \tau'\, \langle n|\partial_{\tau'}|m\rangle \alpha_m(\tau')\nonumber\\
&&~~~~~~~~~~~\times\exp\left[-\int^0_{\tau'} d\tau''\, \big(\mathcal E_n(\tau'')-\mathcal E_m(\tau'') \big) \right],
\label{int_eq}
\eeq
where it should be noted that $\alpha_n(0)=a_n(0)$. In principle we should supply this equation with initial conditions at $\tau=\tau_0$, 
but this is not necessary if $|\tau_0|$ is sufficiently large, since the sensitivity to the initial condition will then be exponentially suppressed. 
Instead, we can impose the asymptotic condition $\alpha_n(\tau\to-\infty)\to \delta_{n0}$, which implies that in the distant past the system was 
in its ground state.

Eq.~(\ref{int_eq}) is ideally suited for an analysis with the APT. In particular, if the rate of change is very small, $\dot\lambda(\tau)\to 0$, 
then to leading order in $\dot\lambda$ the system remains in its ground state; $\alpha_m(\tau)\approx \delta_{m0}$ (except during the initial 
transient, which is not important because we are interested in large $|\tau_0|$). In the next higher order, the transition amplitudes to the states 
$n\neq 0$ are given by;
\be
\alpha_n(0)\approx -\int\limits^0_{-\infty} d\tau \, \langle n|\partial_{\tau}|0\rangle \exp\left[-\int^0_{\tau} d\tau'\, \Delta_{n0}(\tau')\right],
\label{eq_central1}
\ee
where $\Delta_{n0}(\tau)=\mathcal E_n(\tau)-\mathcal E_0(\tau)$.  The matrix element above for non-degenerate states can also 
be written as
\be
\langle n|\partial_\tau|0\rangle =-\langle n|\partial_\tau \mathcal H(\tau)|0\rangle/ \Delta_{n0}(\tau).
\ee
In what follows we will assume that we are dealing with a non-degenerate ground state.

To make further progress in analyzing the transition amplitudes Eq.~(\ref{eq_central1}), we consider the very slow asymptotic limit $\dot\lambda\to 0$. 
To be specific, we assume that near $\tau=0$ the tuning parameter has the form (see also Ref.~[\onlinecite{PhysRevB.81.224301}])
\be
\lambda(\tau)\approx \lambda(0)+\frac{v_\lambda |\tau^r|}{r!} \Theta(-\tau).
\label{vrdef}
\ee
The parameter $v_\lambda$, which controls the adiabaticity, plays the role of the quench amplitude if $r=0$, the velocity for $r=1$, the acceleration for $r=2$, etc. 
It is easy to check that in the asymptotic limit $v_\lambda\to 0$, Eq.~(\ref{eq_central1}) gives
\be
\alpha_n\approx v_\lambda {\langle n|\partial_\lambda|0\rangle \over (\mathcal E_n-\mathcal E_0)^r}=- 
v_\lambda {\langle n|\partial_\lambda \mathcal H|0\rangle \over (\mathcal E_n-\mathcal E_0)^{r+1}},
\label{an1}
\ee
where all matrix elements and energies are evaluated at $\tau=0$. From this perturbative result we can in principle evaluate the leading non-adiabatic 
response of various observables and define the corresponding susceptibilities. For the purposes of comparing with the QAQMC approach, Eq.~(\ref{an1}) suffices.

\subsection{Operator-product evolution} 
\label{sub:qaqmc}

The quasi-adiabatic QMC method may appear very different from NEQMC but has a similar underlying idea. Instead of imaginary time propagation with Eq.~(\ref{utau}), 
we apply a simple operator product to evolve the initial state. We first examine the state propagated with the first $t$ operators in the sequence $P_{t,1}$ 
in Eq.~(\ref{p1m}),
\be
|\psi_t \rangle=[-\mathcal{H}(\lambda_t)] \dots [-\mathcal{H}(\lambda_2)] [-\mathcal{H}(\lambda_1)] |\psi_0\rangle,
\label{psitm}
\ee
and after that we will consider symmetric expectation values of the standard form $\langle \psi_M |A|\psi_M\rangle$ as well as the asymmetric expectation 
values in Eq.~(\ref{atdef}). We assume that the spectrum of $-\mathcal H$ is strictly positive, which is accomplished with a suitable constant offset to 
$\mathcal{H}$ if needed.

\subsubsection{Linear protocols}

The coupling $\lambda$ can depend on the index $t$ in an arbitrary way. It is convenient to define
\be
\tau_i={i\over T},
\ee
where $T$ is the overall time scale, which can be set to unity. The leading non-adiabatic corrections will be determined by the system properties and by 
the behavior of $\lambda(\tau_i)$ near the point of measurement $t$. The most generic is the linear dependence 
$\lambda(\tau_i)\approx \lambda(t)+\tilde v_\lambda (t-\tau_i)$, where $\tilde v_\lambda$ is related to the quench velocity (see below). In the end of 
this section we will briefly consider also more general nonlinear quench protocols.

Our strategy to analyze Eq.~(\ref{psitm}) in the adiabatic limit will be the same as in the preceding subsection. We first go to the instantaneous 
basis and rewrite

\be
|\psi(\tau_i)\rangle\equiv |\psi_i\rangle=\sum_n a_n (\tau_{i}) | n(\lambda_{i})\rangle\equiv \sum_n a_n^i |n^i\rangle .
\ee

In the instantaneous basis, the discrete Schr\"odinger-like equation $|\psi^{i+1}\rangle=-\mathcal H(\tau_{i+1}) |\psi^i\rangle$ reads
\be
a_n^{i+1}=-\sum_m a_m^i \mathcal E_n^{i+1} \langle n^{i+1} | m^{i}\rangle,
\ee
and it is instructive to compare this with Eq.~(\ref{eq 2}). It is convenient to first make a transformation
\be
a_n^i=\prod_{j=i+1}^{t} {1\over (-\mathcal E_n^j)} \alpha_n^i.
\ee
This transformation does not affect the transition amplitude at the time of measurement $t$:  $a_n^t=\alpha_n^t$. Then the equation above becomes
\be
\alpha_n^{i+1}=\sum_m \alpha_m^i \left[\prod_{j=i+1}^t {\mathcal E_n^j\over \mathcal E_m^j}\right] \langle n^{i+1} | m^{i}\rangle.
\ee
Let us introduce a discrete derivative
\be
\langle n^i|\overleftarrow{\Delta} \equiv \langle n^{i+1}|-\langle n^i|,
\ee
and write the Schr\"odinger-like equation as 
\be
\alpha_n^{i+1}=\alpha_n^i+\sum_m \alpha_m^i \left[\prod_{j=i+1}^t {\mathcal E_n^j\over \mathcal E_m^j}\right] \langle n^{i} |\overleftarrow{\Delta}| m^{i}\rangle.
\ee
In the adiabatic limit, the solution of this equation is $\alpha_n^i=\delta_{n0}$, i.e., the instantaneous ground state. To leading order of deviations from 
adiabaticity we find
\be
\alpha_n^{i+1}=C_n+\sum_{k=0}^i\left[\prod_{j=k+1}^t{\mathcal E_n^j\over \mathcal E_0^j}\right]\langle n^{k}|\overleftarrow{\Delta} | 0^{k}\rangle,
\ee
where $C_n$ can be determined from the initial condition. In the limit of sufficiently large $t$ the initial state is not important so we 
should have $\alpha_n^{t-i}\to 0$ for $i\gg 1$, so that $C_n=0$. Therefore we find that the amplitude of the transition to the excited state is approximately
\be
\alpha_n^{t}\approx \sum_{k=0}^{t-1} \left[\prod_{j=k+1}^t{\mathcal E_n^j\over \mathcal E_0^j}\right]\langle n^{k}|\overleftarrow{\Delta} | 0^{k}\rangle .
\ee
Changing the summation index $k$ to $p=t-k$ we have
\be
\alpha_n^{t}\approx \sum_{p=1}^{t} \left[\prod_{j=t+1-p}^t{\mathcal E_n^j\over \mathcal E_0^j}\right]\langle n^{t-p}|\overleftarrow{\Delta} | 0^{t-p}\rangle.
\label{alpha_n}
\ee
It is clear that for large $t$ only $p\ll t$ terms contribute to the sum. In the extreme adiabatic limit one can thus move the matrix 
element outside of the summation and use the spectrum of the final Hamiltonian. In this case we find
\begin{eqnarray}
\alpha_n^t &\approx& {\mathcal E_n\over \mathcal E_0} {\langle n|\overleftarrow{\Delta} | 0\rangle\over 1-{\mathcal E_n/\mathcal E_0}} \nonumber \\ 
           &=&  {-\mathcal E_n\Delta_\lambda} {\langle n|\overleftarrow{\partial_\lambda} | 0\rangle\over \mathcal E_n-\mathcal E_0} 
           = {\mathcal E_n\Delta_\lambda} {\langle n|\partial_\lambda | 0\rangle\over \mathcal E_n-\mathcal E_0},
\label{an2}
\end{eqnarray}
where $\Delta_\lambda=\lambda(t)-\lambda(t-1)$. By comparing Eqs.~(\ref{an1}) and (\ref{an2}) we see that near the adiabatic limit QAQMC and  NEQMC are very 
similar if ${\mathcal E_n/\mathcal E_0}\approx {\rm const}$. This can in principle always be ensured by having a sufficiently large energy offset, but even with 
a small offset we expect the ratio to be essentially constant for the range of $n$ contributing significantly when the spectrum becomes gapless 
close to a quantum-critical point. If the condition indeed is properly satisfied, then from Eqs.~(\ref{an1}) and (\ref{an2}), we identify the quench velocity 
as
\be
v_\lambda=\mathcal E_0\Delta_\lambda. 
\ee
This is the main result of this section. We will confirm its validity explicitly in numerical studies with the QAQMC method in Sec.~\ref{sec:results}. 
Since $\mathcal E_0 \propto N$, where $N$ is the system size, we can also see that $v_\lambda \propto N\Delta_\lambda \propto N/M$ for a given total change 
in $\lambda$ over the $M$ operators in the product.

Let us point out that Eq.~(\ref{an2}) can be also rewritten as
\be
\alpha_n^t \approx -\mathcal E_0\Delta_\lambda {\langle n|\overleftarrow{\partial_\lambda} | 0\rangle\over \mathcal E_n-\mathcal E_0}
-\Delta_\lambda \langle n|\overleftarrow{\partial_\lambda} | 0\rangle.
\label{an3}
\ee
The first contribution here exactly matches that of Eq.~(\ref{an1}) while the second term is an additional contribution corresponding 
to a sudden quench. 

\subsubsection{Nonlinear protocols}

We can extend the above result, Eq.~(\ref{an3}), to arbitrary quench protocols. In particular, consider 
\be
\lambda_{t-p}=\lambda_t+{v_\lambda\over (-\mathcal E_0)^r} {p^{r-1}\over (r-1)!},
\ee
where $r \geq 0$ (not necessarily an integer). For $r=1$, we recover the linear protocol analyzed above. Then we can still rely on Eq.~(\ref{alpha_n}) but need to 
take into account that
\begin{eqnarray}
\langle n^{t-p}|\overleftarrow{\Delta} | 0^{t-p}\rangle &\approx& \Delta\lambda_{t-p}\langle n^t|\overleftarrow{\partial_\lambda}|0^t\rangle \nonumber \\
&=&{v_\lambda\over (-\mathcal E_0)^r} {p^{r-1}\over (r-1)!}
\langle n^t|\overleftarrow{\partial_\lambda}|0^t\rangle.
\end{eqnarray}
Thus, we find that
\be
\alpha_n^t\approx {v_\lambda\over (-\mathcal E_0)^r (r-1)!}{\rm Li_{1-r}(\mathcal E_n/\mathcal E_0)}\langle n^t|\overleftarrow{\partial_\lambda}|0^t\rangle,
\ee
where ${\rm {Li}_{1-r}(x)}$ is the Polylog function. In particular,
\begin{mathletters}
\begin{eqnarray}
&& {\rm Li}_0(x)={1\over 1-x}, \\ 
&& {\rm Li}_{-1}(x)={x\over (1-x)^2},\\ 
&& {\rm Li}_{-2}(x)={x(x+1)\over (1-x)^3}.
\end{eqnarray}
\end{mathletters}
Under the conditions discussed above (large offset or small energy gap) we again have $x=\mathcal E_n/\mathcal E_0\to 1$ and then we recover the continuum  result using the fact that
\[
{\rm Li}_{1-r}(1-\epsilon)\approx {(r-1)!\over \epsilon^r}.
\]
Then, indeed,
\begin{eqnarray}
\alpha_n^t =a_n^t &\approx& {v_\lambda\over (- \mathcal E_0)^r (r-1)!} 
{(r-1)!\over (1-\mathcal E_n/\mathcal E_0)^r}\langle n^t|\overleftarrow{\partial_\lambda}|0^t\rangle \nonumber \\
&=& v_\lambda {\langle n^t|\overleftarrow{\partial_\lambda}|0^t\rangle\over (\mathcal E_n-\mathcal E_0)^r},
\end{eqnarray}
which exactly matches Eq.~(\ref{an1}).

\subsection{Expectation values}
\label{sec:expectation_values}

While asymptotically Eq.~(\ref{atdef}) gives the ground state of the observable $A$ in the adiabatic limit for all values of $t$, the approach to 
this limit as $t\to\infty$ is qualitatively different depending on whether $t$ is equal to $M$ or not. More precisely, if $t=\eta M$ where $\eta\in (0,2)$ as $M\to\infty$, we encounter two different asymptotic regimes for $\eta\neq 1$ and $\eta=1$.

\subsubsection{Symmetric expectation values; $t=M$}

In this limit the expectation value of the observable $A$ in the leading order of the adiabatic perturbation theory reduces to 
\be
\langle A\rangle_{t=M}\approx \langle \psi(v_\lambda) | A|\psi(v_\lambda)\rangle,
\ee
where $v_\lambda\approx \mathcal E_0\Delta_\lambda$ is the imaginary time velocity identified earlier. For generic observables not commuting with the Hamiltonian, we find
\be
\langle A \rangle_{t=M} \approx \langle A\rangle_{0}+v_\lambda \chi'_{A\lambda},
\ee
where 
\be
\chi'_{A\lambda}=\sum_{n\neq 0} \langle 0|A|n\rangle {\langle n| \partial_\lambda|0\rangle\over \mathcal E_n-\mathcal E_0} + c.c.
\label{asusc}
\ee
is the susceptibility. All energies and matrix elements are evaluated at ``time'' $t=M$. 

For diagonal observables $A$, like the energy or energy fluctuations, we have
\be
\langle A\rangle_{t=M}\approx  \langle A\rangle_{0}+v_\lambda^2 \sum_{n\neq 0} 
 {|\langle n| \partial_\lambda|0\rangle|^2\over (\mathcal E_n-\mathcal E_0)^2}\langle n|A|n\rangle.
\label{diag_sym}
\ee
In particular, the correction to the energy is always positive as it should be for any choice of wave function deviating from the ground state.  Let us emphasize that for diagonal observables the leading non-adiabatic response at the symmetric point in imaginary time coincides with that in real time,  and, thus QAQMC or NEQMC can be used to analyze real time deviations from adiabaticity, as was pointed out in the case of NEQMC in Ref.~[\onlinecite{degrandi11}].

\subsubsection{Asymmetric expectation value, $t \neq M$}
\label{subsub:asymmetric}

It turns out that the asymptotic approach to the adiabatic limit is quite different for non-symmetric points $t=\eta M$ with $\eta \neq 1$. Without loss of generality we can focus on $0<\eta<1$ (since all expectation values are symmetric with respect to $\eta \to 2-\eta$  for the symmetric protocol we consider \cite{averagenote}). Then the expectation value of $A$ is evaluated with respect to different eigenstates
\be
\langle A\rangle_t={\langle \psi_L| A|\psi_R\rangle\over \langle \psi_L|\psi_R\rangle},
\ee
where 
\begin{eqnarray}
|\psi_R\rangle&=& \mathcal H(\lambda_t) \cdots \mathcal H(\lambda_2)\mathcal H(\lambda_1) |\psi_0\rangle,\nonumber \\
|\psi_L\rangle&=&\mathcal H(\lambda_{t+1}) \cdots \mathcal H(\lambda_{M-1})\mathcal H(\lambda_M) P_ {M,1} | \psi_0\rangle.
\label{psi_LR}
\end{eqnarray}
Note that the overlap $\langle \psi_L|\psi_R\rangle$ is independent of $t$ by construction and is real.

It is easy to see that for diagonal observables we obtain a leading asymptotic as in Eq.~(\ref{diag_sym}) but with the opposite sign in the second term
\be
\langle A\rangle_{t \neq M}\approx  \langle A\rangle_{0}-v_\lambda^2 \sum_{n\neq 0} 
 {|\langle n| \partial_\lambda|0\rangle|^2\over (\mathcal E_n-\mathcal E_0)^2}\langle n|A|n\rangle.
\label{eq:asy}
\ee
In particular, the leading correction to the ground state energy is negative when $t$ deviates sufficiently from the symmetric point, i.e., 
$  | \lambda_t - \lambda_1| / v_\lambda \ll M $. There is no contradiction here since the left and right states are different (i.e., we are not 
evaluating a true expectation value and there is no variational principle). Both Eqs.~(\ref{diag_sym}) and (\ref{eq:asy}) recover the exact result in the adiabatic limit. Since the correction up to the sign exactly matches the real time result, we can still use the non-symmetric 
expectation value for diagonal observables to extract the real time non-adiabatic response. For $t \to M$, the sign of the correction should 
change, to connect smoothly to the variational $t=M$ expectation value. The crossover between positive and negative corrections to the energy  
takes place around a point that asymptotically converges to $t=M$ in the adiabatic limit (where the deviation from the ground-state energy 
at $t=M$ vanishes). We will illustrate this with numerical results in Sec.~\ref{ecrossover} (see Fig.~\ref{fig1}).

As in the symmetric case, using the APT discussed in the previous section the results derived here easily extend to other 
values of the exponent $r$.

\subsubsection{The metric tensor and Berry curvature}

If $A=-\partial_\mu H$, then the susceptibility Eq.~(\ref{asusc}) reduces to the $\mu\lambda$ component of the metric 
tensor,\cite{degrandi11,adi_long} which, thus, can be readily extracted using the QAQMC algorithm. In particular, the diagonal components of the 
metric tensor define the more familiar fidelity susceptibility.

Next, we observe that for $t$ sufficiently different from $M$, the approach to the ground state in the left function $\psi_L$ in Eq.~(\ref{psi_LR}) 
effectively corresponds to a change in sign of the velocity, and, thus, we find
\be
\langle A \rangle_t\approx {\langle \psi(-v)|A|\psi(v)\rangle \over \langle \psi(-v)|\psi(v)\rangle},
\label{eq:vmv}
\ee

\noindent
where the wave functions $|\psi(v)\rangle$ and $\psi(-v)\rangle$ are evaluated at the same value of the coupling determined by the value of $\eta$.  We can use the results of the previous section to find that for off-diagonal observables
\be
\langle A\rangle_t\approx \langle A\rangle_0-i v_\lambda \chi''_{A\lambda},
\label{eq:off}
\ee

\be
\chi''_{A\lambda}=i\sum_{n\neq 0} \langle 0|A|n\rangle {\langle n| \partial_\lambda|0\rangle\over \mathcal E_n-\mathcal E_0} - c.c.
\ee
Based on this result we conclude that the leading non-adiabatic correction is imaginary and coincides, up to the factor of imaginary $i$, with the real-time 
non-adiabatic correction.\cite{adi_long} In particular, for $A=-\partial_\mu H$ the susceptibility $\chi''_{A\lambda}=\chi''_{\mu\lambda}$ is proportional to 
the Berry curvature. 

The fact that we are getting the opposite sign (compared to the real time protocol)
in the susceptibility for diagonal observables and the Berry curvature for
off-diagonal observables away from the symmetric points in  Eqs~(\ref{eq:asy}) and (\ref{eq:off}) is a
consequence of general analytic properties of the asymmetric expectation values.
As we discuss in Ref.~[\onlinecite{adi_long}] the expectation value Eq.(\ref{eq:vmv}) is the
analytic continuation of the real time expectation value to the imaginary velocity
$v\to iv$. This continuation is valid in all orders of expansion of the expectation
value of $A$ in $v$.

\section{Results}
\label{sec:results}

As a demonstration of the utility of QAQMC and the behaviors derived in the previous section we here study the TFIM, 
defined by the Hamiltonian
\begin{equation}
   \label{eq:hamiltonian}
   \mathcal{H} =  - s  \sum_{\langle i,j \rangle}  \hspace{2pt}  \sigma_{i}^z   \hspace{1pt} \sigma_{j}^z  - (1-s) \sum_{i} \sigma_i^x,
\end{equation}
\noindent
where $\langle i,j \rangle$ are nearest-neighbor sites, and $\sigma_z$ and $\sigma_x$ are Pauli matrices. Here, $s$ plays the role of the 
tuning parameter, which in the simulations reported below will vary between $0$ (where the ground state is trivial) to a value exceeding the
quantum-critical point; $s_c=1/2$ in a 1D chain and $s_c \approx 0.247$ in the 2D square lattice.\cite{J.Phys.A.33.6683}

We work in the standard basis of eigenstates of all $S^z_i$. The simulation algorithm samples strings of $2M$ diagonal and off-diagonal
terms in Eq.~(\ref{eq:hamiltonian}), in a way very similar to the $T>0$ stochastic series expansion (SSE) method, which has been discussed in detail in
the case of the TFIM in Ref.~[\onlinecite{PhysRevE.68.056701}]. The modifications for the QAQMC primarily concern the sampling of the initial state,
here $|\Psi(0)\rangle = \prod_{i}|\uparrow_i + \downarrow_i\rangle$, which essentially amounts to a particular boundary condition replacing the
periodic boundaries in finite-temperature simulations. An SSE-like scheme with such modified boundaries was also implemented for the NEQMC
method in Ref.~[\onlinecite{degrandi11}], and recently also in a study of combinatorial optimization problems in Ref.~[\onlinecite{farhi12}]. We here 
follow the same scheme, using cluster updates in which clusters can be terminated at the boundaries. The implementation for the product with varying 
coupling $s$ is even simpler than SSE or NEQMC, with the fixed-length product replacing the series expansion of Eq.~(\ref{utau}). The changes 
relative to Refs.~[\onlinecite{degrandi11}] and [\onlinecite{PhysRevE.68.056701}]  are straightforward and we therefore do not discuss the sampling 
scheme further here.

\subsection{Cross-over of the energy correction}
\label{ecrossover}

\begin{figure}
   \includegraphics[width=7cm, clip=true]{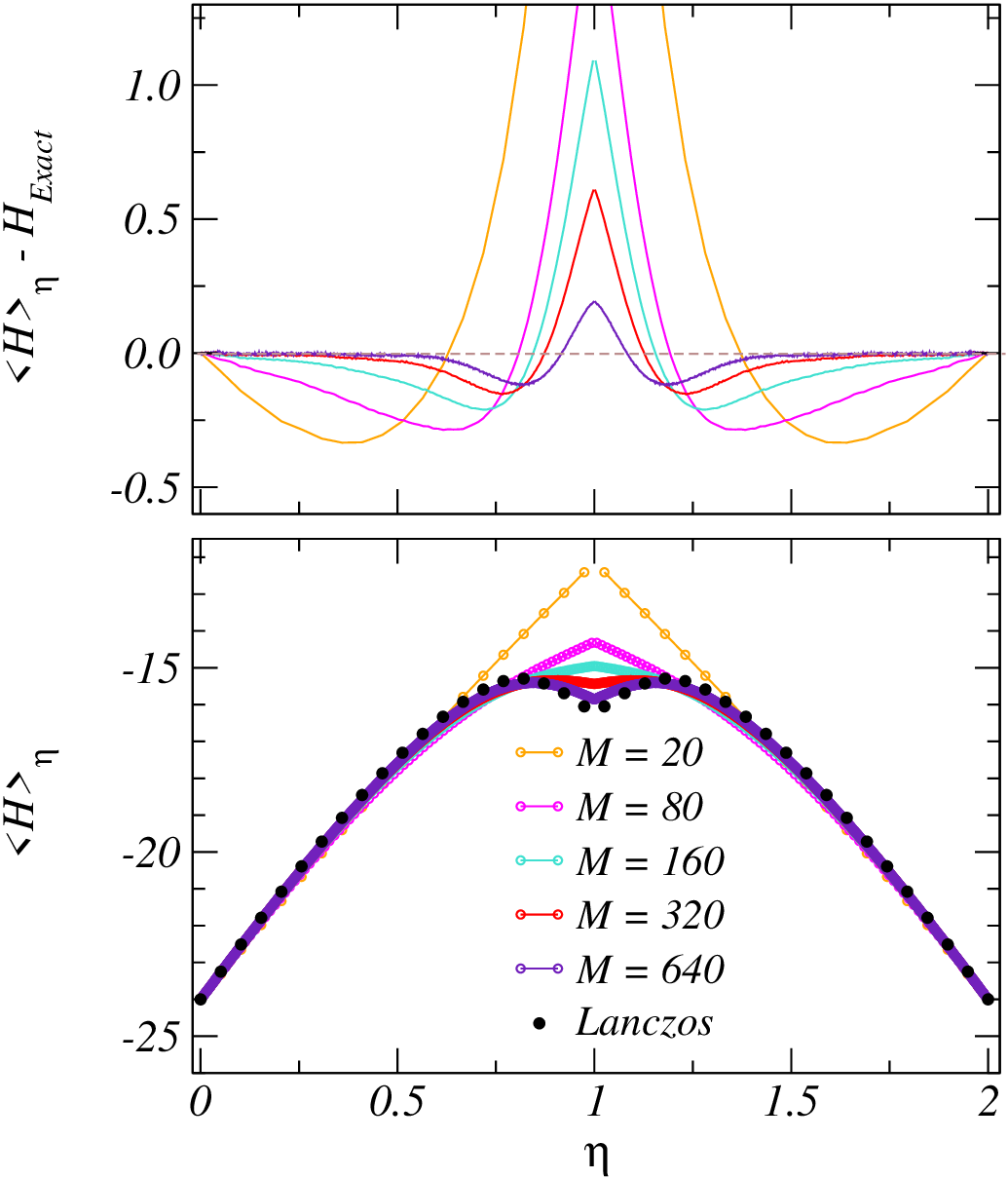}
\caption{(Color online) Symmetric and asymmetric expectation values of the Hamiltonian in QAQMC calculations for 1D TFIM Eq.~(\ref{eq:hamiltonian})
with $N=24$. Here, the evolution was from $s=0$ to $0.6$ and, thus, $s=0.6$ is the symmetric point here labeled by $\eta=t/2M=1$. For $\eta\le 1$,
$s=0.6\eta$ and for $\eta\ge 1$, $s=1.2-0.6\eta$, and the critical point $s=1/2$ hence corresponds to $\eta_{c1} \approx 0.833$  and $\eta_{c2} \approx 1.167$. 
(Bottom) Expectation value and (top) deviation from the true ground-state energy (obtained using Lanczos exact diagonalization).}
\label{fig1}
\end{figure}

As we discussed in Sec.~(\ref{sec:apt}), the asymmetric expectation value (\ref{atdef}) of the Hamiltonian has a negative correction to the ground-state energy when 
$t$ is sufficiently away from the symmetric point $t=M$. In Fig.~\ref{fig1} we illustrate this property and the convergence to the ground-state energy for
all $t$ with increasing $M$ with simulation data for a small 1D TFIM system. We here plot the results versus the rescaled propagation power $\eta=t/M$. 
The region of negative deviations move toward the symmetric point with increasing $M$. Note that the deviations here are not strongly influenced by the 
critical point (which is within the parameter $s$ simulated but away from the symmetric point), although the rate of convergence should also be slow due to criticality. 
The rate of convergence to the ground state can be expected to be (and is here seen to be) most rapid for $\eta < \eta_{c1}$ and $\eta > \eta_{c2}$.

\subsection{Quantum-critical dynamic scaling}

The idea of dynamic scaling at a critical point dates back to Kibble and Zurek for quenches (also called ramps, since the parameter does not have to 
change suddenly, but linearly with arbitrary velocity as a function of time) of systems through classical phase transitions. \cite{J.Phys.A.9.1387,Nature.317.505}
Here, the focus was on the density of defects. The ideas were later generalized also to quantities more easily accessible in experiments, such as order
parameters, and the scaling arguments were also extended to quantum systems.\cite{PhysRevB.72.161201,RevModPhys.83.863,PhysRevLett.95.105701,PhysRevLett.95.245701} 
The basic notion is that the system has a relaxation time $t_{\rm rel}$, and if some parameter (here a parameter of the Hamiltonian) is changed such that a critical 
point is approached, the system can stay adiabatic (or in equilibrium) only if the remaining time $t$ to reach the critical point is much larger than the 
relaxation time, $t \gg t_{\rm rel}$. In general, one expects $t_{\rm rel} \sim \xi^z \sim \epsilon^{-z\nu}$, where $\xi$ is the correlation length, $\nu$ the 
exponent governing its divergence with the distance $\epsilon$ to the critical point, and $z$ the dynamic exponent. For a system of finite size (length) 
$L$, $\xi$ is maximally of order $L$ and, thus, for a linear quench the critical velocity $v_{\rm crit}$ separating slow and fast power-law quenches according 
to Eq.~(\ref{vrdef}) should heuristically be given by $v_{crit} \sim L^{-(z + 1/\nu)}$, and for a power-law quench with exponent $r$ according to Eq.~(\ref{vrdef}) 
this generalizes to \cite{PhysRevB.81.224301} 
\begin{equation}
   \label{eq:critical_v}
    v_{crit} \sim L^{-(zr + 1/\nu)}.    
\end{equation}
One then also expects a generalized finite-size scaling form for singular quantities $A$,
\begin{equation}    
 A(L,\epsilon)  = L^\kappa f(\epsilon L^{1/\nu}, v L^{zr+ 1/\nu}),
\label{eq:universal_f}
 \end{equation}
where $\kappa$ characterizes the leading size-dependence at the critical point of the quantity considered. 
For $v \to 0$, Eq.~(\ref{eq:universal_f}) reduces to the standard equilibrium finite-size scaling hypothesis. This scaling was recently 
suggested and tested in different systems, both quantum~\cite{deng_08,PhysRevB.81.224301, kolodrubetz_12} and classical~\cite{chandran_12}.
 
The above expression Eq.~(\ref{eq:universal_f}) combined with the product-evolution Eq.~(\ref{p1m}) allows us to study a phase transition based on different
combinations of scaling in the system size and the velocity in non-equilibrium setups. For example, if one wants to find the critical point for the phase 
transition and the exponent $\nu$ is known, one can carry out the evolution under the critical-velocity condition:
 \begin{equation}
    \label{eq:const}
    v L^{z+1/\nu} = c,
 \end{equation}
where $c$ is a constant. In this paper, we focus on linear quench protocols and set $r=1$ henceforth. As we discussed in Sec.~\ref{sub:qaqmc}, the QAQMC 
method applied to a system of size (volume) $N$ based on evolution with $M$ operators in the sequence and change $\Delta_\lambda$ between each successive 
operator corresponds to a velocity $v \propto N \Delta_\lambda \propto N/M$, with the prefactor depending on the ground state energy (at the critical point). 
The exact prefactor will not be important for the calculations reported below, and for convenience in this section, we define
\be
v = s_{\rm f} \frac{N}{M},
\label{vnmdef}
\ee
where $s_{\rm f}$ is the final value of the parameter $s$ in Eq.~(\ref{eq:hamiltonian}) over the evolution (which is also the total change in $s$, 
since we start with the eigenstate at $s=0$). The critical product-length $M$ is, thus, given by
\be
M = \frac{1}{c}NL^{z+1/\nu} =  \frac{1}{c}L^{d+z+1/\nu},
\label{mscaled}
\ee
where we have also for simplicity absorbed $s_{\rm f}$ into $c$.

Using an arbitrary $c$ of order $1$ in Eq.~(\ref{eq:const}), the critical point $s_c$ can be obtained  based on a scaling function with the single 
argument $\epsilon L^{1/\nu}$ in Eq.~(\ref{eq:universal_f}). We will test this approach here, in Secs.~\ref{sub1d} and \ref{sub2d}, and later, 
in Sec.~\ref{subfurther}, we will show that exact knowledge of the exponents in Eq.~(\ref{eq:const}) is actually not needed. First, we discuss
the quantities we consider in these studies.

\subsection{Quantities studied}

We will focus our studies here on the squared $z$-component magnetization (order parameter),
 \begin{equation}
   \label{eq:m2}
   m_z^2 = \Big\langle \frac{1}{N^2} \bigg( \sum_i^N \sigma_i^z \bigg)^2 \Big\rangle ,
\end{equation}
We can also  define a susceptibility-like quantity (which we will henceforth refer to as the susceptibility) measuring 
the magnetization fluctuations:
\begin{equation}
   \label{eq:susceptibilty}   
    \chi = N ( \left\langle m_z^2 \right\rangle -   \left\langle |m_z| \right\rangle^2).
\end{equation}
\noindent
Here both terms have the same critical size-scaling as the equal-time correlation function;
\be
\label{eq:msquare}
\langle m_z\rangle ^2 \sim  \left\langle |m_z| \right\rangle^2 \sim L^{-(d+z-2+\eta)}, 
\ee
where $d$ is the spatial dimensionality. The prefactors for the two quantities are different, however, a divergent peak remains in Eq.~(\ref{eq:susceptibilty}) 
at the transition. Away from the critical point $\chi \to 0$ with increasing system size.

To clarify our use of $\chi$, we point out that we could also just study the scaling of $\left\langle m_z^2 \right\rangle$, but
the peak produced when subtracting off the second term in Eq.~(\ref{eq:susceptibilty}) is helpful in the scaling analysis. According to Eq.~(\ref{eq:universal_f})
and using $z=1$ in Eq.~(\ref{eq:msquare}), the full scaling behavior of the fluctuation around the critical point should follow the form
\begin{equation}
   \label{eq:scaled_susceptibilty}   
    \chi \sim L^{1-\eta} \hspace{2pt}   f \big( (s-s_c) L^{1/\nu}, v L^{1 + 1/\nu } \big),
\end{equation}
for any dimensionality $d$. 

We should point out here that the true thermodynamic susceptibility based on the Kubo formula \cite{PhysRevB.43.5950} 
(imaginary-time integral) yields a stronger divergence $L^{2-\eta}$. This quantity is, however, more difficult to study with
the QAQMC algorithm, because, unlike in standard finite-$T$ QMC methods, the time integration cannot simply be carried out within the space of time-evolving 
Hamiltonians in Eq.~(\ref{p1m}) and Eq.~(\ref{atdef}). The standard Feynman-Suzuki correspondence between the $d$-dimensional quantum and $(d+1)$-dimensional 
classical systems is not realized in our scheme. The configuration space of time-evolving Hamiltonians builds in the relaxation time, $t_{\rm rel}$, in a 
different way, not just in terms of equilibrium fluctuations in the time direction, but in terms of evolution as a function of a time-dependent parameter. 

A useful quantity to consider for extracting the critical point is the Binder cumulant, \cite{PhysRevLett.47.693},
  \begin{equation}
    \label{eq:binder}
   U = \frac{3}{2} \bigg( 1 - \frac{1}{3} \dfrac{ \left\langle m_z^4 \right\rangle }{ \left\langle m_z^2 \right\rangle^2 } \bigg).
 \end{equation}
For a continuous phase transition, $U$ converges to a step function as $L \to \infty$. The standard way to analyze this quantity for finite 
$L$ is to graph it versus the argument $s$ for different $L$ and extract crossing points, which approach the critical
point with increasing $L$. Normally, this is done in the equilibrium, either by taking the limit of the temperature $T \to 0$ for each $L$ first, or
by fixing $\beta=1/T \propto L^z$ if $z$ is known. Here, the latter condition is replaced by Eq.~(\ref{eq:const}), but, as we will discuss further below,
the condition can be relaxed and the exponents do not have to be known accurately \textit{a priori}. Our approach can also be used to determine the exponents, 
either in a combined procedure of simultaneously determining the critical point and the exponents, or with a simpler analysis after first determining 
the critical point.

We have up until now only considered calculations of equal-time observables, but, in principle, it is also possible and interesting to study correlations 
in the evolution direction, which also can be used to define susceptibilities.

In the following we will illustrate various scaling procedures using results for the 1D and 2D TFIMs. The dynamic exponent $z=1$ is known for both cases, 
and in the 1D case all the exponents are rigorously known since they coincide with those of the classical 2D Ising model. For the 2D TFIM, the exponents are
know rather accurately based on numerics for the 3D classical model.
 
\subsection{1D transverse-field Ising model.} 
\label{sub1d}

\begin{figure}
   \includegraphics[width=7.5cm, clip=true]{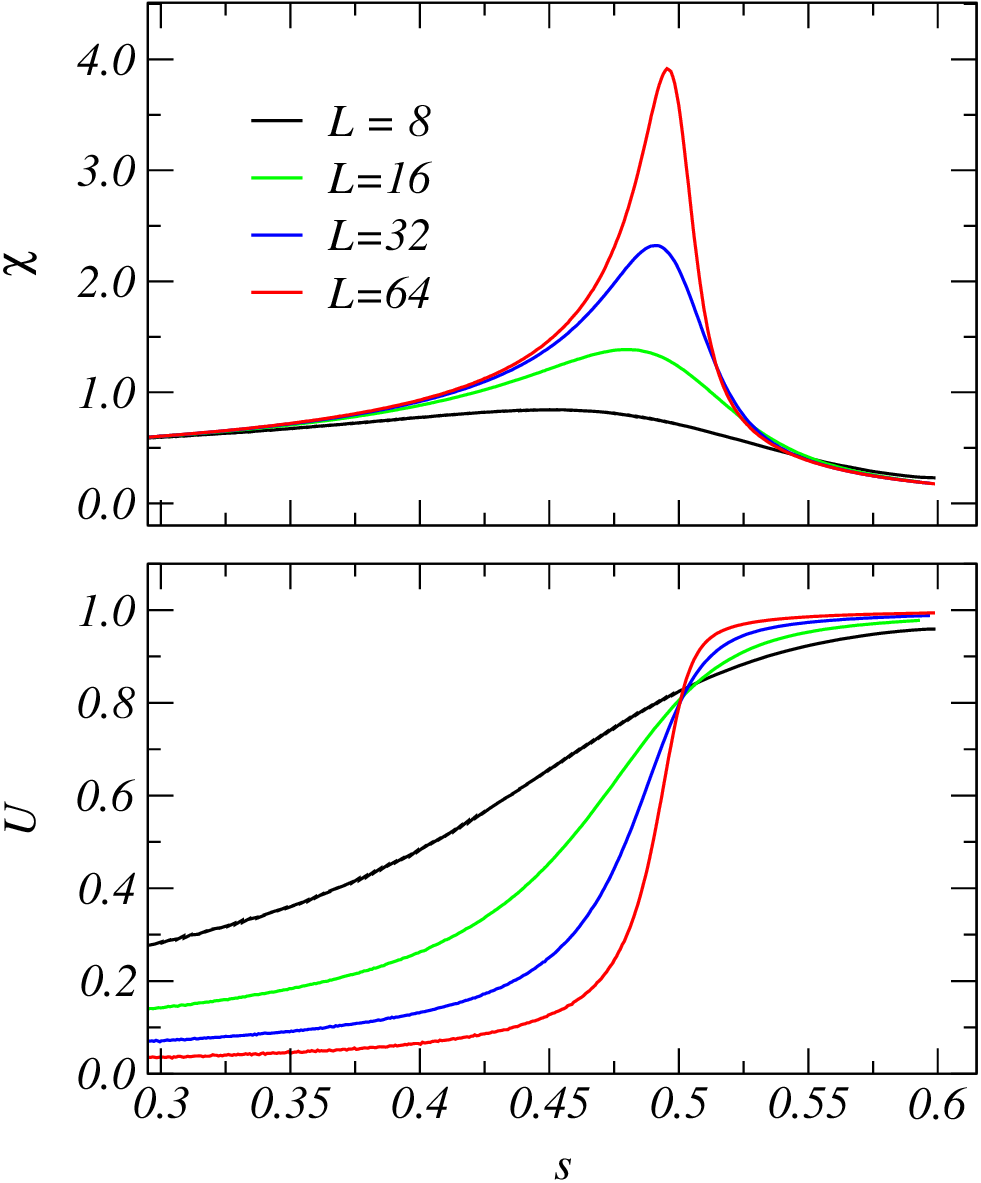}
\caption{(Color online) Results of typical QAQMC runs for the 1D TFIM, Eq.~(\ref{eq:hamiltonian}). The binder cumulant
Eq.~(\ref{eq:binder}) (bottom) and the susceptibility $\chi$ Eq.~(\ref{eq:susceptibilty}) (top) are graphed versus $s$ for several
system sizes $L$. In these simulations, which spanned the range $s \in [0,0.6]$, the length of the index sequence was of the form Eq.~(\ref{mscaled}), 
i.e., with the exponents applicable in this case $M=L^3/c$ with the arbitrary constant chosen to be $c=4^3/240$.}
\label{fig2}
\end{figure}

The 1D TFIM provides a rigorous testing ground for the new algorithm and scaling procedures since it can be solved exactly.\cite{sachdev_qpt} The critical 
point corresponds to the ratio between the transverse field and the spin-spin coupling equaling $1$, i.e., $s=1/2$ in the Hamiltonian Eq.~(\ref{eq:hamiltonian}).
The critical exponents, known through the mapping to the 2D Ising model,\cite{Prog.Theor.Phys.56.1454} are $\nu=1$ and $\eta = 1/4$. 

The results presented here were obtained in simulations with the parameter $s$ spanning the range $[0,s_{\rm f}]$ with $s_{\rm f}=0.6$, i.e., 
going from the trivial ground state of the field term to well above the critical point. Fig.~\ref{fig2} shows examples of results for the susceptibility 
and the Binder cumulant. The operator-sequence length $M$, Eq.~(\ref{p1m}), was scaled with the system size in order to stay at the critical velocity 
according to Eq.~(\ref{mscaled}).
We emphasize again that a single run produces a full curve within the $s$-range used. In order to focus on the behavior close to criticality, we have left out the 
results for small $s$ in Fig.~\ref{fig2}. Since $M$ is very large (up to $\approx 10^6$ for the largest $L$ in the cases shown in the figure), we also do not compute 
expectation values for each $t$ in Eq.~(\ref{atdef}), but typically spacing measurements by $\propto N$ operators.

\begin{figure}
   \includegraphics[width=8cm, clip=true]{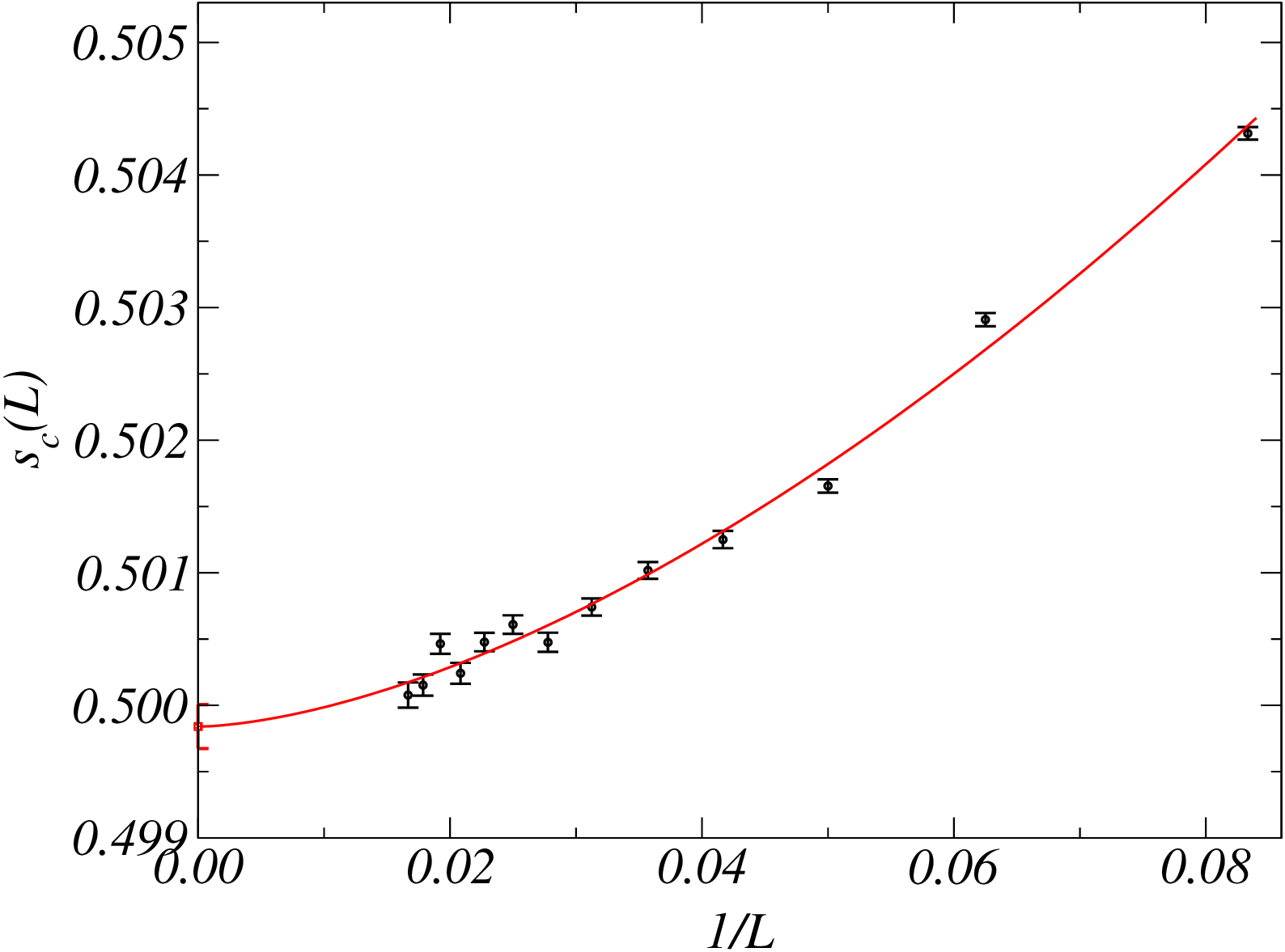}
   \caption{ (Color online) Results of a Binder-crossing scaling analysis of the 1D TFIM data in Fig.~\ref{fig2} (including also other system sizes not shown there). 
             Crossing points were extracted based on system sizes $L$ and $L+4$, with $L=4, 8, \dots, 60$. The curve is a fit to the form \cite{PhysRevLett.47.693} $s_c(L) = s_c + a/L^b$, 
             $s_c=0.49984(16)$ and $b=1.6(1)$. }
   \label{fig3}
\end{figure}

Extracting Binder curve-crossings using system-size pairs $L$ and $L+4$, with $L=4, 8, 12, \dots 60$, and extrapolating the results to $L \rightarrow \infty$, 
we find $s_c = 0.49984(16)$, as illustrated in Fig.(\ref{fig3}). Thus, the procedure produces results in full agreement with the known critical point.

\begin{figure}
   \includegraphics[width=7.5cm, clip=true]{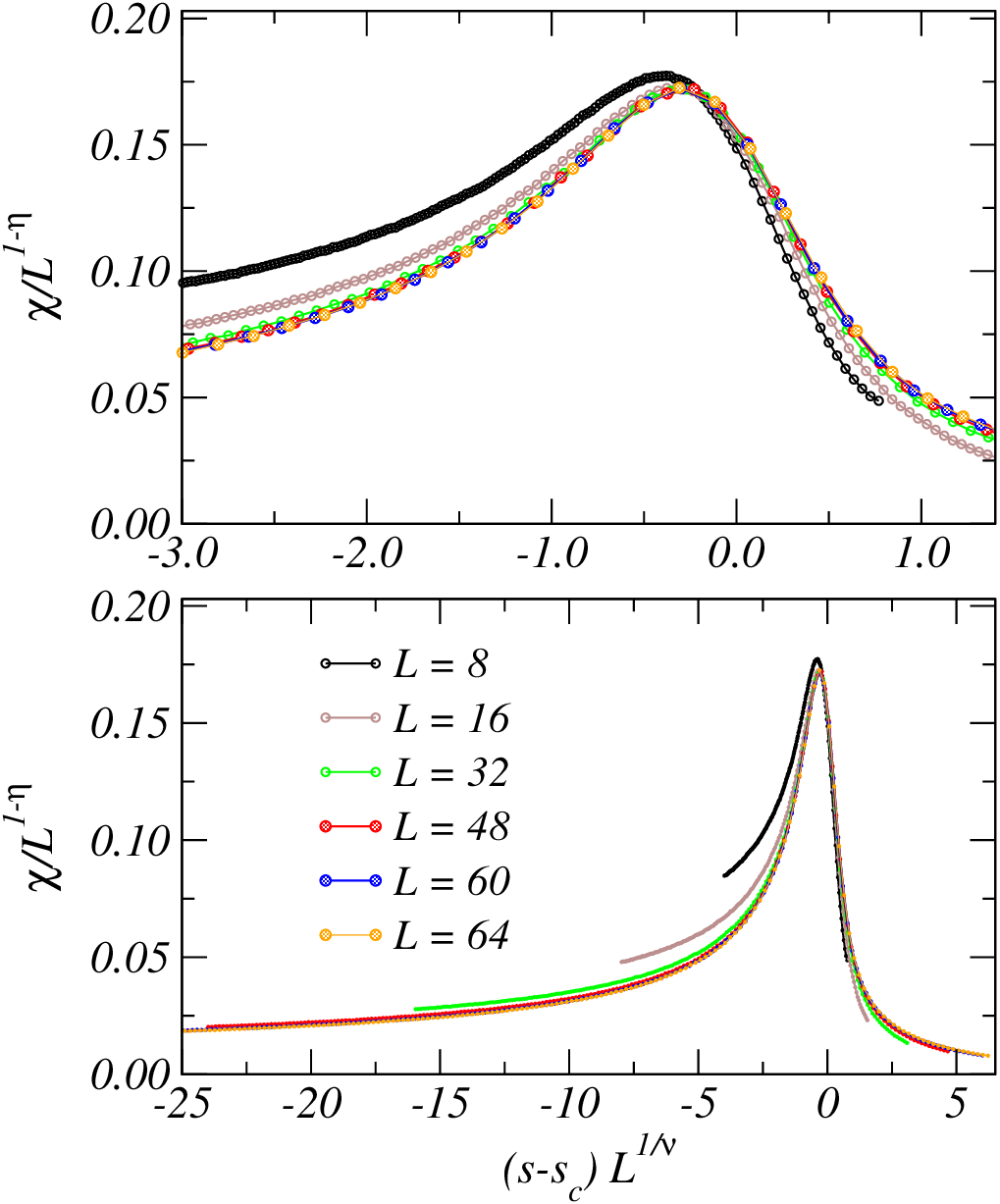}
   \caption{ (Color online) Scaled susceptibility of the 1D TFIM. The axes have been scaled according to the form Eq.~(\ref{eq:scaled_susceptibilty}) with the second argument constant and using the exact critical point $s_c=1/2$. The results are shown on two different scales to make visible deviations (due to subleading size and velocity corrections) from the common scaling function far away from criticality as well as the good          data collapse close to the critical point.}
   \label{fig4}
\end{figure}

The dynamical scaling of the susceptibility is illustrated in Fig.~\ref{fig4}. Here, there are no adjustable parameters at all, since all exponents
and the critical coupling are known (and we use the exact critical coupling $s_c=1/2$, although the numerical result extracted below is very close to this
value and produces an almost identical scaling collapse). While some deviations from a common scaling function are seen for the smaller systems and
far away from the scaled critical point $(s-s_c)L$, the results for larger sizes and close to the peak rapidly approach a common scaling function.
This behavior confirms in practice our discussion of the definition of the velocity and the ability of the QAQMC method to correctly take into account  at least the first corrections to the adiabatic evolution.

\subsection{2D transverse-field Ising model}
\label{sub2d}

The 2D transverse-field Ising model provides a more serious test for our algorithm since it cannot be solved exactly. Among many previous numerical 
studies,\cite{J.Phys.C.4.2370, PhysRevB.57.8494,J.Phys.A.33.6683} Ref.~[\onlinecite{J.Phys.A.33.6683}] arguably has the highest precision so far for the
value of the critical coupling ratio. Exact diagonalization was there carried out for up to $6 \times 6$ lattice size. In terms of the critical 
field $h_c=1-s$ in units of the coupling $J=s$, the critical point was determined to $h_c/J = 1/0.32841(2) = 3.04497(18)$, where the error bar reflects 
estimated uncertainties in finite-size extrapolations. Results based on QMC simulations \cite{J.Phys.C.4.2370, PhysRevB.57.8494} are in agreement with 
this value, but the statistical errors are larger than the above extrapolation uncertainty. One might worry that the system sizes $L\le 6$ are very small 
and the extrapolations may not reflect the true asymptotic $L\to \infty$ size behavior. However, the data points do follow the functional forms expected based 
on the corresponding low-energy field theory, and there is therefore no \textit{a priory} reason to question the results. It is still useful to try to reach similar 
or higher precision with other approaches, as we will do here with the QAQMC method combined with dynamic scaling.

\begin{figure}
   \includegraphics[width=7.5cm, clip=true]{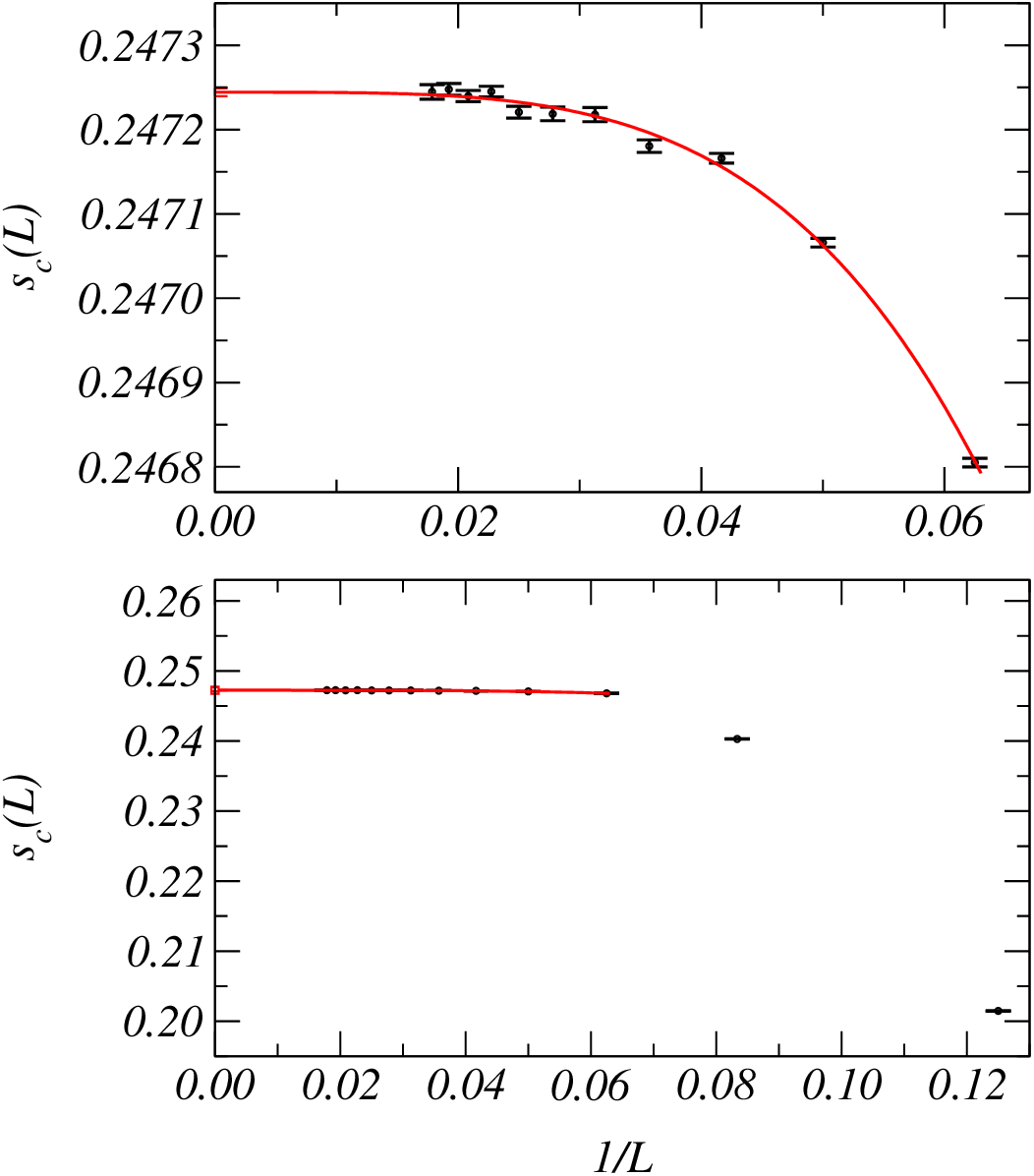}
   \caption{(Color online) Binder crossings for the 2D TFIM extracted using $L$ and $L+4$ systems with  $L=4, 8, \dots, 56$. 
           The crossing points have been fitted to the standard form \cite{PhysRevLett.47.693} $s_c(L) = s_c + a/L^b$, for which the optimal values are $s_c=0.247244(4)$
           and $b=4.0(1)$. The results are shown on two different scales to illustrate large deviations from the fitted form for the smaller systems, 
           followed by a rapid convergence for larger sizes.}
   \label{fig5}
\end{figure}

In this case we simulate the linear quench in the window of $s \in [0,0.3]$, which contains the previous estimates for the critical value $s_c \approx 0.247$ as discussed above. Although we could also carry out an independent scaling analysis to extract the critical exponents, we here choose to just use their values based on previous work on the classical 3D Ising model; $1/\nu \approx 1.59$, and $\eta \approx 0.036$.\cite{PhysRevB.59.11471} Our goal here is to extract a high-precision estimate of the critical coupling, and, at the same time, to further test the ability of QAQMC to capture the correct critical scaling behavior. We again scale $M$ with $L$ according to Eq.~(\ref{mscaled}), with the constant $c = 4^{4.59}/32$.

As in the 1D case, we extract Binder-cumulant crossing points based on linear system sizes $L$ and $L+4$ with $L = 4, 8, \dots, 56$. Fig.~\ref{fig5} shows the results versus $1/L$ along with a fit to a power-law correction \cite{PhysRevLett.47.693} for $s_c(L)$. Extrapolating to infinite size gives $s_c = 0.247244(4)$, which corresponds to a critical field (in unit of $J$) $h_c/J = 3.04458(7)$. This is in reasonable good agreement with the value obtained in Ref.~[\onlinecite{J.Phys.A.33.6683}] and quoted above, with our (statistical) error bar being somewhat smaller. To our knowledge, this is the most precise value for the critical coupling of this model obtained to date. We emphasize that we here relied on the non-equilibrium scaling ansatz to extract the equilibrium critical point. Allowing for deviations from adiabaticity in a controlled way and utilizing the advantage of the QAQMC algorithm allowed us to extract observables in the whole range of couplings in a single run. This requires considerably less computational resources than standard equilibrium simulations, which must be repeated for several different couplings in order to carry out the crossing-point analysis.

\begin{figure}
   \includegraphics[width=7.5cm, clip=true]{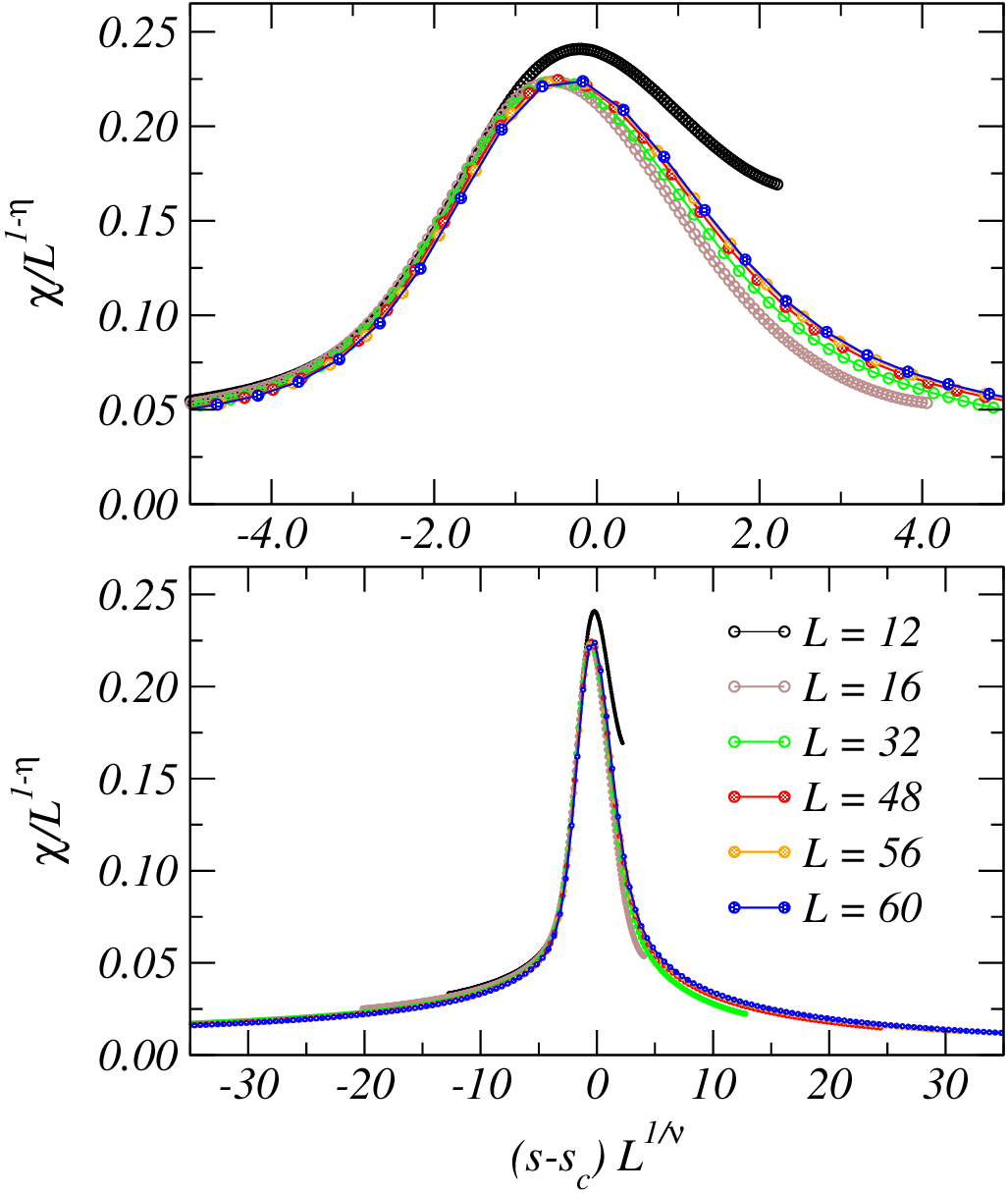}
   \caption{(Color online) Scaled susceptibility of the 2D TFIM, based on Eq.~(\ref{eq:scaled_susceptibilty}) with a constant second argument.
            Here we have used $1/\nu = 1.59$ and $\eta = 0.036$ for the classical 3D Ising model \cite{PhysRevB.59.11471}.}
   \label{fig6}
\end{figure}

Fig.~\ref{fig6} shows the susceptibility scaled according to the behavior expected with Eq.~(\ref{eq:universal_f}) when the second argument is held
constant. As in the 1D case, the data converge rapidly with increasing size toward a common scaling function in the neighborhood of
the transition point, again confirming the correct quasi-adiabatic nature of the QAQMC method.

\subsection{Further tests}
\label{subfurther}

The results discussed in the preceding subsections were obtained with the KZ velocity condition Eq.~(\ref{eq:const}), applied in the form of Eq.~(\ref{mscaled})
tailored to the QAQMC approach, with specific values for the constant  $c$. In principle, the constant is arbitrary, but the non-universal details of the 
scaling behavior depend on it. This is in analogy with a dependence on the shape, e.g., an aspect ratio, of a system in equilibrium simulations at finite 
temperature, or to the way the inverse temperature $\beta=1/T$ is scaled as $aL^z$ with arbitrary $a$ in studies of quantum phase transitions (as an alternative
to taking the limit $\beta \to \infty$ for each lattice size). The critical point and the critical exponents should not depend on the choices of such shape factors
or limiting procedures.

To extract the critical coupling,
in the preceding subsections, we fixed the exponents $\nu$ and $z$ at their (approximately) known values, 
and one may at first sight assume that it is necessary to use their correct values. It is certainly
some times convenient to do so, in order to set the second argument of the scaling function Eq.~(\ref{eq:universal_f}) to a constant and, thus, obtain a simpler 
scaling function  depending on a single argument. However, one can study critical properties based on the scaling approach discussed above as long as the 
velocity approaches zero as the system size increases. This observation can be important in cases where the critical exponents are not known and one would like 
to obtain an accurate estimate of the critical coupling before carrying out a scaling analysis to study exponents. We will test this in practice here. As we 
will discuss further below, one should use a different power $\kappa$ in the scaling ansatz Eq.~(\ref{eq:universal_f}) if the velocity is brought to zero slower 
than the critical form.

In cases where we use the ``wrong'' values of the exponents, we formally replace $z+1/\nu$ by a free parameter $\alpha$, 
\be
v \sim L^{-\alpha}/c,
\label{valphascaled}
\ee
 and
the corresponding substitution in Eq.~(\ref{mscaled}). To understand the scaling of the observables for arbitrary $\alpha$, we return to the general scaling 
form given by Eq.~(\ref{eq:universal_f}). In the case of the Binder cumulant and for linear quench protocol, this form reads
\begin{equation}    
U  = f \big((s-s_c) L^{1/\nu}, v L^{z+1/\nu}\big).
\label{eq:universal_g}
 \end{equation}
As we pointed out above, when the velocity scales exactly as $L^{-(z+1/\nu)}$, the dependence on the second argument in the scaling function drops out and we 
can find the crossing point in a standard way as we did in Figs.~\ref{fig3} and \ref{fig5}. Suppose that we do not know the exponents 
$\nu$ and $z$ \textit{a priory} and instead scale $v$ as in Eq.~(\ref{valphascaled}). Then there are three possible situations: (i) $\alpha=z+1/\nu$, 
(ii) $\alpha>z+1/\nu$, and (iii) $\alpha<z+1/\nu$, where we already have analyzed scenario (i). In scenario (ii), where velocity scales to zero 
faster than the critical KZ velocity, the second argument of the scaling function $L^{z+1/\nu}/L^{\alpha}$ approaches zero as the system size increases and,
thus, the scaling function effectively approaches the equilibrium velocity-independent form. We can then extract the crossing point as in the first scenario, 
and this gives the correct critical coupling in the limit of large system sizes. Finally, in case (iii) the velocity scales zero slower than the critical 
KZ value and the second argument in Eq.~(\ref{eq:universal_g}) diverges, which implies that the system enters a strongly non-equilibrium regime. This 
scenario effectively corresponds to taking the thermodynamic limit first and the adiabatic limit second. Then, if the system is initially on the disordered 
side of the transition, the Binder cumulant vanishes in the thermodynamic limit. At finite but large system sizes its approach to zero should be given 
by the standard Gaussian theory:
\be
U\approx {C\over L^d}.
\label{binder_therm}
\ee
Combining this with the scaling ansatz Eq.~(\ref{eq:universal_g}) we find that for $\alpha<z+1/\nu$, the expected asymptotic of the Binder cumulant is
\be
U\approx L^{-d}v^{-{d}/{(z+1/\nu)}} \tilde f \big((s-s_c) L^{1/\nu} \big),
\label{rescaled_binder}
\ee
where $\tilde f$ is some other velocity independent scaling function. Thus we can find the correct transition point by finding crossing points of 
$UL^d v^{d/(z+1/\nu)}$. Similar considerations apply to the ordered side of the transition, where the Binder cumulant approaches one as the 
inverse volume.

\begin{figure}
   \includegraphics[width=8cm]{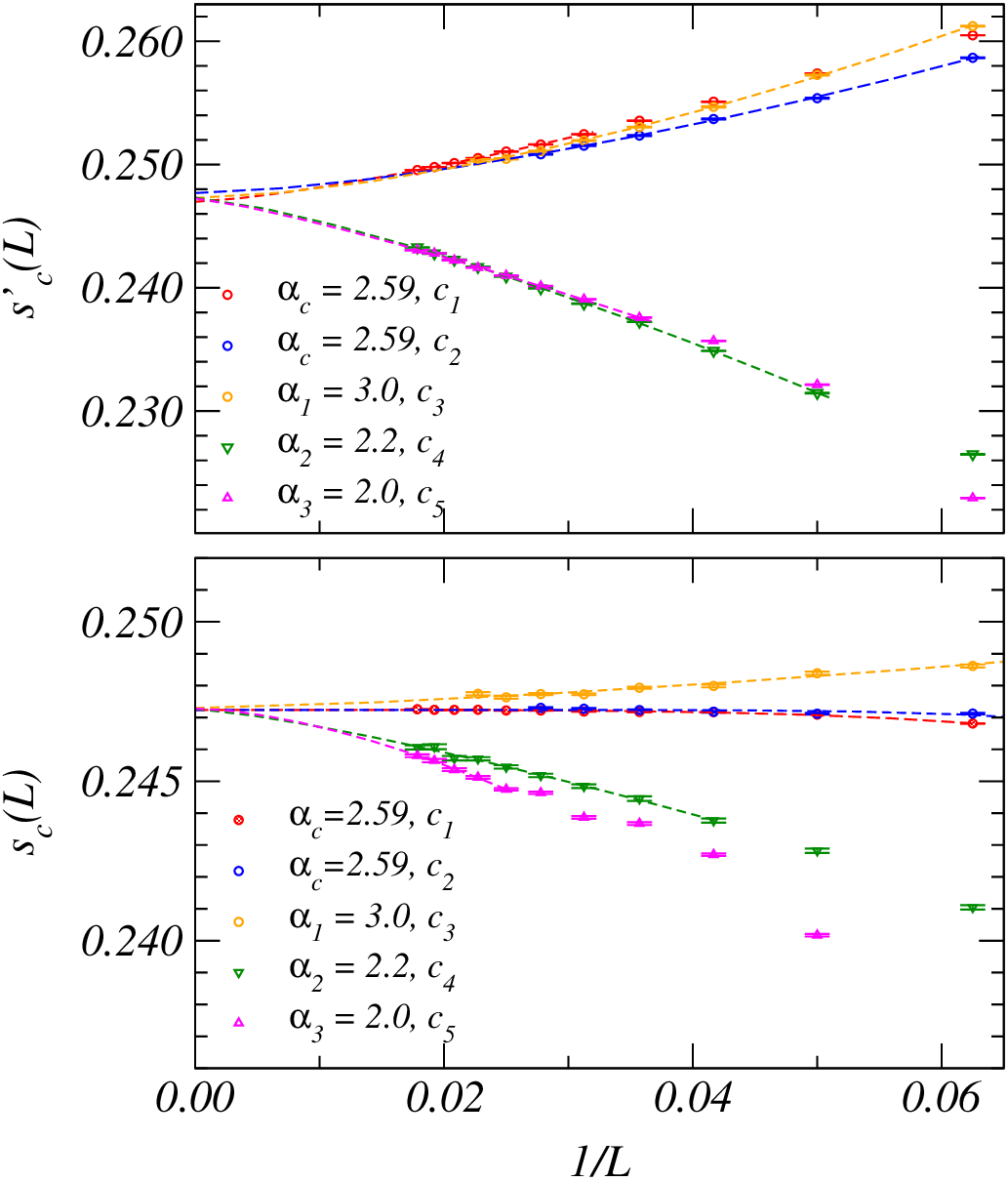}
   \caption{ (Color online) Critical-point estimates based on curve crossings of appropriately scaled quantities for scenarios (i)-(iii)
             discussed in the text. The Binder cumulant (bottom) and the squared magnetization (top) give estimates $s_c(L)$ and 
             $s'_c(L)$, respectively, based on system sizes $L$ and $L+4$. The red and blue curves correspond to runs in which the velocity was kept at 
             the critical value, scenario (i), but with different constants of proportionality $c$ in Eq.~(\ref{mscaled}); $c_1 = 4^{4.59}/32$ 
             and $c_2 = 4^{4.59}/48$. The yellow curves were obtained with the velocity decreasing faster than $v_{crit}$ with $L$, scenario (ii), 
             with the proportionality constant $c_3= 4^5/32$. The green and pink curves correspond to cases where the velocity is sub-critical, 
             scenario (iii), with constants $c_4= 4^{4.2}/32$, $c_5 = 4^4/32$. In all cases, power-law corrections were fit in order to extrapolate 
             to infinite size (with small sizes excluded until statistically sound fits were obtained).}
   \label{fig7}
\end{figure}

The three cases are illustrated in the lower panel of Fig.~\ref{fig7}, which shows Binder-cumulant crossings extracted from appropriately scaled data
in cases (i), (ii), and (iii) above. Additionally, to illustrate the insensitivity to the choice of the constant $c$ in the scaled sequence length 
in Eq.~(\ref{mscaled}), results based on two different constants are shown for case (i). In all cases, the extrapolated critical couplings agree with 
each other to within statistical errors.  Note that, on the one hand, if the exponent $\alpha$ gets very large, then the time of simulations, 
which scales as $M$, rapidly increases with the system size and the algorithm becomes inefficient. On the other hand, if $\alpha$ is very small,
our results indicate that the size dependence is larger and it is more difficult to carry out the extrapolation to infinite size. The optimal value 
of $\alpha$ should be as close as possible to the critical KZ power, but to be on the safe side when scaling according to the standard KZ critical
form, case (i), one may choose a somewhat larger value, since the subcritical velocity in case (ii) has the same scaling form.

Next we illustrate how the same idea works in the case of the order parameter. Around the critical point ($s_c$, $v_{\text{crit}}$), the squared
magnetization [see Eq.~(\ref{eq:m2})] can be written as
\begin{eqnarray}
m_z^2 & = & L^{-2\beta/\nu} f \big(  (s- s_c) L^{1/\nu}, v L^{z+1/\nu} \big).
\label{mz2}
\end{eqnarray}
As in the previous discussion we scale $v\sim L^{-\alpha}$ and depending on the exponent $\alpha$ there are two different asymptotics of 
the scaling function. For $\alpha\geq z+1/\nu$ the second argument vanishes or approaches constant so we effectively get the equilibrium scaling  
\begin{eqnarray}
m_z^2 & = & L^{-2\beta/\nu} f \big(  (s- s_c) L^{1/\nu}\big) 
\end{eqnarray}
If, conversely, $\alpha<z+1/\nu$ then on the disordered side of the transition $m_z^2$ scales as $L^{-d}$. This immediately determines the asymptotic 
of the scaling function in Eq.~(\ref{mz2}):
\begin{equation}
\label{eq:rescaled_m2}
m_z^2 = L^{-d} v^{(2\beta/\nu)-d \over z+1/\nu} \tilde f\big(  (s- s_c) L^{1/\nu}\big).
\end{equation}
Equation (\ref{eq:rescaled_m2}) can be used in the same way as the Binder cumulant to extrapolate the critical point, using the standard 
form \cite{PhysRevLett.47.693} $s'_c(L) = s'_c + a/L^b$
for the rescaled $m_z^2$. As shown in the top panel of Fig.~(\ref{fig7}), after rescaling the order parameter and extrapolating the crossing points between 
the appropriately rescaled $m_z^2$ curves to the thermodynamics limit, all curves, obtained from below or above the adiabatic limit Eq.~(\ref{eq:critical_v}), 
converge to the same value $s'_c \approx 0.247$. This approach also suggests a way to determine the transition point in experiment, since one can sweep 
through the critical point at different velocities, the crossing point can then be extracted through the measurement of the order parameter. It is also 
worth mentioning that since one can extrapolate the critical point independently without knowing the actual exponent $\nu$ prior to the simulation, 
an optimization procedure can be carried out to determine the exponents posterior to the simulation.\cite{classical_quench.unpublished}

\begin{figure}
   \includegraphics[width=7.5cm, clip=true]{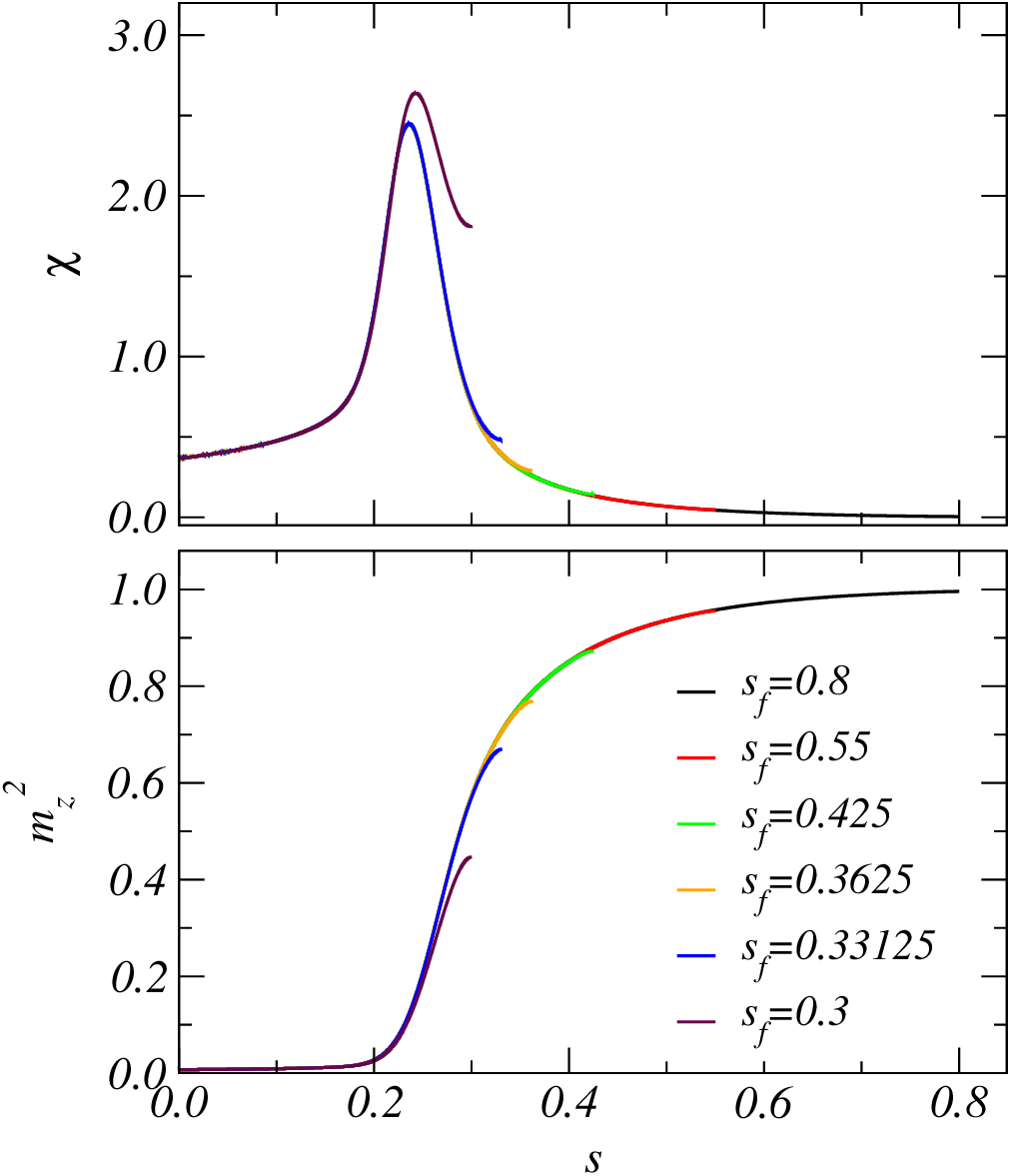}
   \caption{(Color online) Squared magnetization (bottom) and susceptibility (top) vs $s$ of the 2D TFIM with $L=12$. In these runs, different curves correspond to different end points $s_f$ of the evolution, with the velocity $v \propto s_{\rm f}N/M$ kept constant. The $s_f=0.3$ curve is from the simulation shown in Sec.~\ref{sub2d}.}
   \label{fig8}
\end{figure}

For completeness we also briefly discuss the role of the final point $s_{\rm f}$ of the evolution. Fig.~\ref{fig8} shows 2D results for the squared magnetization Eq.~(\ref{eq:m2}) and susceptibility Eq.~(\ref{eq:susceptibilty}) obtained for a range of final points above the critical value. Here the velocity was kept constant for all the cases. The values of the computed quantities at some fixed $s$, e.g., at $s_c$, show a weak dependence on $s_{\rm f}$ for the lowest-$s_{\rm f}$ runs. The 
deviations are caused by contributions of order $v^2$ and higher, which are non-universal as discussed in Sec.~\ref{sub:qaqmc}. For very high velocities 
the dependence on $s_{\rm f}$ can be much more dramatic than in Fig.~\ref{fig8}, but this is not the regime in which the QAQMC should be applied to study 
universal physics.

\section{Summary and Discussion}
\label{sec:conclusions}

We have presented a nonequilibrium QAQMC approach to study quantum dynamics, with a simple product of operators with evolving coupling replacing the
standard Schr\"odinger time evolution. We showed that this approach captures the leading non-adiabatic corrections to the adiabatic limit, both by
analytical calculations based on the APT and by explicit simulations of quantum-critical systems with the QAQMC algorithm. The simulation results obey the
expected generalized dynamic scaling with known static and dynamic critical exponents. We also extended the scaling formalism beyond results obtained 
previously in Ref.~[\onlinecite{degrandi11}]. We analyzed the leading non-adiabatic corrections within this method and showed that they can be used to 
extract various non-equal time correlation functions, in particular, the Berry curvature and the components of the metric tensor. A clear advantage of 
the QAQMC approach is that one can access the whole range of couplings in a single run. Being a simple modification of projector QMC, the QAQMC method 
is applicable to the same class of models as this conventional class of QMC schemes---essentially models for which ``sign problems'' can be avoided.

As an illustration of the utility of QAQMC, we applied the algorithm and the scaling
procedures to the 1D and 2D TFIMs. The expected scaling behaviors are observed very clearly. In the 1D case we extracted
a critical coupling in full agreement with the known value, and in 2D we obtained an estimate with unprecedented (to our knowledge) precision (small
error bars); $(h/J)_c=3.04458(7)$. Based on repeating the fitting procedures with different subsets of the data, we believe that any systematical errors 
due to subleading corrections neglected in the extrapolations should be much smaller than the statistical errors, and, thus, we consider the above 
result as unbiased.

The QAQMC approach bears some similarities to previous implementations of {\it quantum annealing} within QMC algorithms.\cite{Santoro06,Bapst12} However, 
the previous works have mainly considered standard equilibrium QMC approaches in which some system parameter is changed as a function of the 
{\it simulation time}. This evolution is not directly related to true quantum dynamics (and, thus, is not really quantum annealing), but is dependent 
on the particular method used to update the configurations. In contrast, in our scheme, as in the NEQMC method introduced in Ref.~[\onlinecite{degrandi11}], 
the evolution takes place {\it within} the individual configurations, and there is a direct relationship to true Schr\"odinger evolution in imaginary time. 

In Green's function (GF) QMC simulations the gradual change of a system parameter with the simulation time is rather closely related to the QAQMC scheme 
(since also there one applies a series of terms of the Hamiltonian to a state), with the difference being that QAQMC uses true importance sampling of configurations,
with no need for guiding wave functions and no potential problems related to mixed estimators. Our asymmetric expectation values could be considered
as a kind of mixed estimator as well, but we have completely characterized them within the APT. In addition, the previous uses of GFQMC with
time-evolving Hamiltonians have, to our knowledge, never addressed the exact meaning of the velocity of the parameter evolution. The correct definition of 
the velocity is of paramount importance when applying quantum-critical scaling methods, as we have discussed here. We have here computed the
velocity within APT for the QAQMC scheme. The same relationship with Schr\"odinger dynamics may possibly hold for GFQMC as well, but, we have 
not applied the APT to this case and it is therefore not yet clear whether GFQMC can capture correctly the same universal non-equilibrium susceptibilities 
as the QAQMC and NEQMC methods. We expect QAQMC to be superior to time-evolving  GFQMC, because of its better control over measured symmetric and asymmetric 
expectation values and fully realized importance sampling.  

Some variants of GFQMC use true importance sampling, e.g., the Reptation QMC (RQMC) method,\cite{PhysRevLett.82.4745} which also avoids mixed estimators. 
The configuration space and sampling in the QAQMC method bears some similarities with RQMC, recent lattice versions of which also use SSE-inspired 
updating schemes.\cite{PhysRevE.82.046710} However, to our knowledge, imaginary-time evolving Hamiltonians have not been considered in RQMC and in other 
related variants of GFQMC, nor has the role played by the velocity when crossing the quantum critical point been stressed. This has so far been our
focus in applications of the QAQMC and NEQMC methods. In principle one could also implement the ideas of time-evolution similar to QAQMC within the 
RQMC approach.

We also stress that we have here not focused on optimization. Previous works on quantum annealing within QMC schemes have typically focused
on their abilities to optimize difficult classical problems. While the QAQMC may potentially also offer some opportunities in this direction, our
primary interest in the method is to use it to extract challenging dynamical information under various circumstances. A recent theoretical analysis
of optimization within sign-problem free QMC approaches \cite{Hastings13} is not directly applicable to the QAQMC and NEQMC approaches but 
generalizations should be possible. 

The QAQMC and NEQMC methods 
provide correct realizations of quantum annealing in imaginary time. Besides their ability to study dynamic scaling, with exponents identical to those 
in real-time Schr\"odinger dynamics,\cite{degrandi11} it will be interesting to explore what other aspects of real-time dynamics can be extracted 
with these methods. In particular, their applicability to quantum glasses, of interest in the context of quantum adiabatic computing \cite{farhi12}
as well as in condensed matter physics, deserves further studies.

The ability of the QAQMC to produce results for a whole evolution path in a single run can in principle also be carried over to the conventional
Schr\"odinger imaginary-time evolution with $U(\tau)$ in Eq.~(\ref{utau}). By ``slicing'' the time evolution into $K$ successive evolutions over
a time-segment $\Delta_\tau$,
\begin{equation}
U(\tau)=\prod_{n=1}^K T_{\tau} {\rm exp}\left [ - \int_{\tau_{n-1}}^{\tau_n} d \tau \mathcal{H}[\lambda(\tau)] \right ],
\label{utauk}
\end{equation}
where $\tau_n=n\Delta_\tau$,
one can evaluate matrix elements analogous to Eq.~(\ref{atdef}) by inserting the operator of interest at any  point within the
product of time-slice operators in $\langle \Psi(\lambda_0)|U^*(\tau)U(\tau)|\Psi(\lambda_0)\rangle$. In this case, the symmetric expectation value,
evaluated at the mid-point, is identical to the NEQMC method,\cite{degrandi11} and the asymmetric expectation values will exhibit properties similar to those
discussed in Sec.~\ref{subsub:asymmetric}. We have not yet explored this approach, and it is not clear whether it would have any other advantage besides the exact reduction to Schr\"odinger dynamics of the symmetric expectation values. In practice the simulations will be  more complex than the
QAQMC approach because of the need to sample integrals, but not much more so than the NEQMC method. It should be relatively easy to adapt the RQMC method with an evolving Hamiltonian in this formulation of the time-evolution.

\begin{figure}
   \includegraphics[width=7.5cm, clip=true]{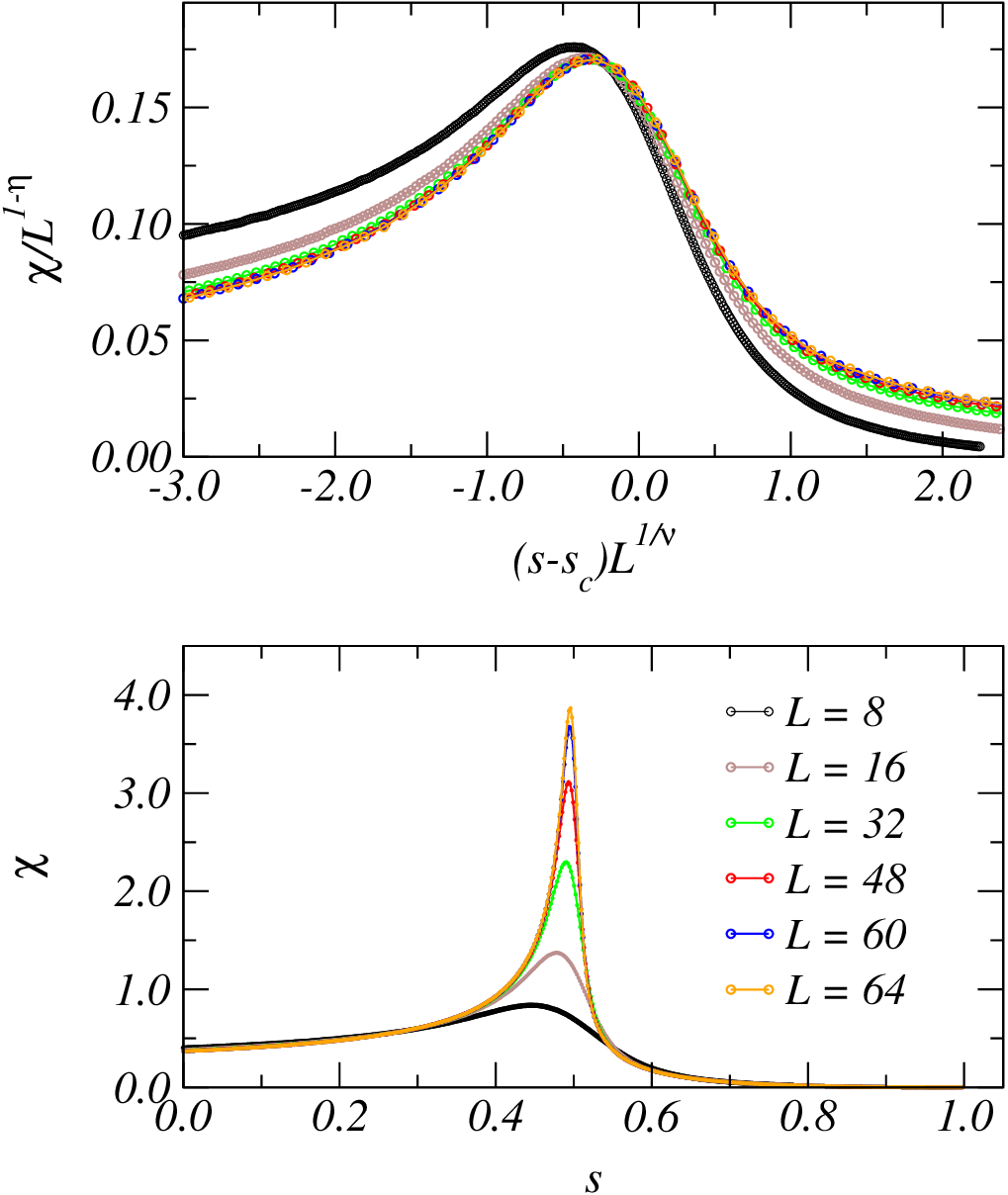}
   \caption{(Color online) One-way evolution $s \in [0,1]$ with QAQMC for the 1D TFIM. (Bottom) The susceptibility Eq.~(\ref{eq:susceptibilty}). 
  (Top) The rescaled susceptibility Eq.~(\ref{eq:scaled_susceptibilty}). Each full curve corresponding to a given chain length $L$ was obtained 
   in a single run. The constant for the critical-velocity condition Eq.~(\ref{eq:const}) was held at $4^3/80$.}
   \label{fig9}
\end{figure}

Finally, we point out that, in principle, one can also carry out a {\it one-way evolutions} with the QAQMC algorithm. Instead of starting with the $\lambda=\lambda_0$ eigenstate at both $\langle \psi_L|$ and $|\psi_R\rangle$ and then projecting them to the $\lambda = \lambda_M$ eigenstate using two sequences of the form Eq.~(\ref{p1m}), one can make $\langle \psi_L|$ correspond to $\lambda_0$ and let it evolve to $|\psi_R\rangle$ corresponding to $\lambda_M$ with only a single operator sequence of length $M$. In the case of the TFIM Eq.~(\ref{eq:hamiltonian}), the obvious choice is then to evolve from $s=0$ to $s=1$ (the classical Ising model), so that both edge states are trivial. All our conclusions regarding the definition of the velocity and applicability of scaling form remain valid in this one-way QAQMC. Results demonstrating this in the case of the 1D TFIM are sown in Fig.~\ref{fig9}. We anticipate that this approach may be better than the two-way evolution in some cases, but we have not yet compared the two approaches extensively.

\begin{acknowledgments}

We acknowledge collaboration with Claudia De Grandi in a related work and would like to thank Mike Kolodrubetz for valuable comments. This work was supported 
by the NSF under Grant No.~PHY-1211284.

\end{acknowledgments}


\begin{thebibliography}{99}

\bibitem{assaad07}
F. F. Assaad and H. G. Evertz, Lect. Notes Phys. {\bf 739}, 277 (2007).

\bibitem{sandvik10}
A.~W. Sandvik,
AIP Conf. Proc. {\bf 1297}, 135 (2010).

\bibitem{kaul12}
R. K. Kaul, R. G. Melko, and A. W. Sandvik, Annu. Rev. Cond. Mat. Phys. {\bf 4}, 179 (2013).

\bibitem{betanote}
Here $\beta \propto M/N$ gives the same rate of convergence for the two choices for a given system size $N$. This can be shown, e.g., by a 
Taylor-expansion of the exponential, which for large $\beta$ is dominated by powers $n \approx \beta |E_0|$, where $E_0$ is the ground 
state energy.

\bibitem{PhysRevLett.69.2863}
S. R. White,
Phys. Rev. Lett. {\bf 69}, 2863 (1992)

\bibitem{RevModPhys.77.259}
U.~Schollw\"ock, Rev. Mod. Phys. {\bf 77}, 259 (2005).

\bibitem{degrandi11}
C. De Grandi, A. Polkovnikov, and A. W. Sandvik,
Phys. Rev. B {\bf 84}, 224303 (2011).

\bibitem{gridnote}
In principle one can also consider a nonlinear grid of ``time'' points, but here we will consider the simplest case of a uniform grid. 

\bibitem{averagenote}
Eq.~(\ref{atdef}) has been defined for $t\le M$, but we can also define the asymmetric expectation value for $2M \ge t>M$ by placing the operator within the product $P_{M,1}$. Clearly, we have the symmetry $\langle A\rangle_{2M-t}=\langle A\rangle_{t}$.

\bibitem{dorneich01}
A. Dorneich and M. Troyer, Phys. Rev. E {\bf 64}, 066701 (2001). 

\bibitem{provost_80}
J.~P. Provost and G.~Vallee, Comm. Math. Phys., {\bf 76}, 289 (1980).

\bibitem{adi_long} C. De Grandi, A. Polkovnikov and A. Sandvik,
{arXiv:1301.2329}.

\bibitem{ortiz_2008}
G.~Rigolin, G.~Ortiz, and V. H.~Ponce,
Phys. Rev. A {\bf 78}, 052508 (2008). 

\bibitem{PhysRevB.81.224301}
C.~De~Grandi, V.~Gritsev, and A.~Polkovnikov,
Phys. Rev. B {\bf 81}, 224301 (2010).

\bibitem{J.Phys.A.33.6683}
C.~J. Hamer,
J. Phys. A: Math. Gen. {\bf 33}, 6683 (2000).

\bibitem{PhysRevE.68.056701}
A.~W. Sandvik,
Phys. Rev. E {\bf 68}, 056701 (2003).

\bibitem{farhi12}
E. Farhi, D. Gosset, I. Hen, A. W. Sandvik, P. Shor, A. P. Young, and F. Zamponi,
Phys. Rev. A {\bf 86}, 052334 (2012).

\bibitem{J.Phys.A.9.1387}
T.~W.~B. Kibble,
J. Phys. A: Math. Gen. {\bf 9}, 1387 (1976).

\bibitem{Nature.317.505}
W.~H. Zurek,
Nature {\bf 317}, 505 (1985).

\bibitem{PhysRevB.72.161201}
A.~Polkovnikov,
Phys. Rev. B {\bf 72}, 161201 (2005).

\bibitem{RevModPhys.83.863}
A.~Polkovnikov, K.~Sengupta,  A. Silva and M.~Vengalattore,
Rev. Mod. Phys. {\bf 83}, 863 (2011)

\bibitem{PhysRevLett.95.105701}
W.~H. Zurek, U.~Dorner, and P.~Zoller,
Phys. Rev. Lett. {\bf 95}, 105701 (2005).

\bibitem{PhysRevLett.95.245701}
J.~Dziarmaga,
Phys. Rev. Lett. {\bf 95}, 245701 (2005).

\bibitem{deng_08}  S.~Deng, G.~Ortiz, L.~Viola,
Europhys. Lett. {\bf 84}, 67008 (2008).

\bibitem{kolodrubetz_12} M. Kolodrubetz, B. K. Clark, and D. A. Huse,  
Phys. Rev. Lett., {\bf 109}, 015701, (2012).

\bibitem{chandran_12} A. Chandran,  A. Erez, S. S. Gubser, S. L. Sondhi,
Phys. Rev. B {\bf 86}, 064304 (2012)

\bibitem{PhysRevB.43.5950}
A.~W. Sandvik and J.~Kurkij\"arvi,
Phys. Rev. B {\bf 43}, 5950 (1991).

\bibitem{PhysRevLett.47.693}
K.~Binder,
Phys. Rev. Lett. {\bf 47}, 693 (1981).

\bibitem{sachdev_qpt}
S. Sachdev, \textit{Quantum Phase Transitions} (Cambridge University Press, Cambridge, 2012) .

\bibitem{Prog.Theor.Phys.56.1454}
M.~Suzuki,
Prog. Theor. Phys. {\bf 56}, 1454 (1976).

\bibitem{J.Phys.C.4.2370}
P.~Pfeuty and R.~J. Elliott,
J. Phys. C {\bf 4}, 2370 (1971).

\bibitem{PhysRevB.57.8494}
M.~S.~L. du~Croo~de Jongh and J.~M.~J. van Leeuwen,
Phys. Rev. B {\bf 57}, 8494 (1998).

\bibitem{PhysRevB.59.11471}
M.~Hasenbusch, K.~Pinn, and S.~Vinti,
Phys. Rev. B {\bf 59}, 11471 (1999).

\bibitem{classical_quench.unpublished}
C.-W. Liu, A.~Polkovnikov, and A.~W. Sandvik (unpublished).

\bibitem{Santoro06}
G. E. Santoro and E. Tosatti, 
J. Phys. A: Math. Gen. {\bf 39} R393 (2006).

\bibitem{Bapst12}
V. Bapst, L. Foini, F. Krzakala, G. Semerjian, and F. Zamponi,
Phys. Rep. {\bf 523} 127 (2013).

\bibitem{PhysRevLett.82.4745}
S. Baroni and S. Moroni,
Phys. Rev. Lett. {\bf 82} 4745 (1999),

\bibitem{PhysRevE.82.046710}
G. Carleo, F. Becca, S. Moroni, and S. Baroni,
Phys. Re. E {\bf 82} 046710 (2010).

\bibitem{Hastings13}
M. B. Hastings, M. H. Freedman, arXiv:1302.5733.

\end{thebibliography}
\end{document}